# Electrical and seismic refraction methods: fundamental concepts, current trends, and emerging machine learning prospects—A review

**Adedibu Sunny Akingboye** [a, b, c]

[a] *Department of Earth Sciences, Adekunle Ajasin University, 001 Akungba-Akoko, Ondo State, Nigeria*

[b] *Geophysics Programme, Universiti Sains Malaysia, 11800 USM, Penang, Malaysia*

[c] *Helmholtz Centre Potsdam – GFZ German Research Centre for Geosciences, 14473 Telegrafenberg, Potsdam, Brandenburg, Germany*

*Correspondence to:* adedibu.akingboye@aaua.edu.ng

**ABSTRACT**

This comprehensive review examines electrical and seismic refraction methods, emphasizing their advanced applications in electrical resistivity tomography (ERT) and seismic refraction tomography (SRT). These techniques are crucial for understanding surface–subsurface crustal dynamics, offering critical insights into resistivity and velocity structures for geological and geohazard assessments. The review also explores the induced polarization (IP) and self-potential (SP) methods as complementary approaches. Despite their effectiveness, ERT and SRT face challenges due to lithological heterogeneities, complex geological processes, and geophysical data uncertainties, necessitating multidisciplinary solutions such as methodological advancements and data integration strategies. Recently, machine learning (ML) techniques have been increasingly applied to joint ERT and SRT analyses, optimizing nonlinear inversion processes and improving the characterization of complex subsurface lithologies. The case studies presented in this review evaluate how supervised and unsupervised ML techniques enhance ERT and SRT by improving data interpretation, refining inversion accuracy, automating lithological differentiation, and predicting seismic velocity from resistivity data. The findings underscore the importance of integrating traditional geophysical methods with advanced data-driven approaches to improve subsurface investigations. Continued innovations in ERT and SRT methodologies, along with emerging computational tools and ML applications, will further enhance their effectiveness in geological, hydrological, environmental, and hazard assessments.

## 1. Introduction

Exploring the intricate dynamics of surface–subsurface crustal formations is fundamental to geoscience research, driving the development of innovative methodologies to uncover hidden complexities of the Earth's subsurface (Abdullah et al. 2022; Akingboye and Bery 2023a; Sujitapan et al. 2024; Akingboye 2024a). Subsurface investigations rely on direct and indirect geophysical methods, which apply physics-based principles to assess subsurface structures, material properties, and geological processes (Loke et al. 2022; Akingboye and Bery 2023b; Hasan et al. 2025; Arif et al. 2025). Given the complexity of geological formations, integrating geophysical, geological, and geochemical data with advanced inversion models enhances imaging. Among geophysical methods, electrical and seismic refraction techniques are essential tools in groundwater assessment, mineral exploration, geotechnical investigations, geoengineering, archaeology, and agrogeophysics (Cheng et al. 2019; Bièvre et al. 2021; Sujitapan et al. 2024). However, conventional approaches often struggle with subsurface heterogeneities, variable geophysical responses, and data uncertainties. Advancing methodologies through improved inversion models and computational techniques is essential for better subsurface characterization (Binley 2015; Dick et al. 2024a; Doyoro et al. 2025).

The development of electrical resistivity tomography (ERT) and seismic refraction tomography (SRT) has significantly improved the resolution and accuracy of subsurface investigations (Zeng et al. 2018; Ronczka et al. 2018; Penta de Peppo et al. 2024; Doyoro et al. 2025). ERT maps electrical resistivity distributions, offering insights into lithology, fluid content, and structural features, governed by mineral composition, porosity, and pore fluid conductivity (Marquis and Hyndman 1992; Meju et al. 2003; Akingboye 2023; Dick et al. 2025). The IP method further refines ERT by analyzing subsurface polarizability for mineralogical differentiation and fluid detection (Kemna et al. 2004; Binley and Kemna 2005; Akingboye and Bery 2021a; Bala et al. 2024; Wang et al. 2025). SRT, leveraging seismic wave velocity variations, delineates geological structures, fracture zones, and subsurface cavities (Dahlin and Wisén 2018; Sujitapan et al. 2024). When integrated with IP and SP methods, geophysical interpretations improve—IP detects clay-rich and water-saturated zones via chargeability contrasts (Slater & Glaser 2003), while SP identifies electrochemical subsurface reactions (Biswas and Sharma 2017; Hasan and Shang 2023).

Accurate subsurface characterization requires evaluating geological and physical properties such as porosity, fracture density, and mineral composition. However, integrating resistivity and seismic datasets presents challenges due to differences in resolution, scale, and sensitivity to geological variations (Bièvre et al. 2021;

Akingboye and Bery 2022; Akingboye 2024b; Dick et al. 2024a). These complexities intensify when merging multiple geophysical datasets that capture distinct physical processes (Gallardo and Meju 2003, 2007). Since subsurface targets are inherently three-dimensional, 3-D ERT and SRT surveys, along with advanced inversion models, provide more reliable solutions for complex subsurface conditions (Zhang et al. 2021; Hasan et al. 2023). Optimizing survey design, inversion techniques, and interpretation methods enhances data accuracy, improves signal-to-noise ratios, and ensures robust geological modeling, ultimately increasing efficiency and reducing operational costs.

With the increasing complexity of joint resistivity and seismic nonlinear inversion, machine learning (ML) techniques have become powerful tools for improving ERT and SRT data processing ML and deep learning (a subset of ML) refine inversion models, automate data interpretation, and enhance prediction accuracy (Yang and Ma 2019; Liu et al. 2020). Traditional geophysical inversions rely on subjective parameter selection and are computationally intensive, limiting their effectiveness in complex geological environments. In contrast, ML-based approaches streamline data processing, enhance pattern recognition, and enable adaptive learning, leading to more efficient geophysical interpretations. These advancements significantly improve resistivity and velocity model resolution, offering deeper insights into subsurface structures.

Supervised and unsupervised ML algorithms—including artificial neural networks (ANNs), convolutional neural networks (CNNs), Monte Carlo methods, support vector machines (SVMs), random forest (RF) classifiers, boosting ensembles (gradient boost [GBoost] and categorical boost [CatBoost]), generative adversarial networks (GANs), and Bayesian models—demonstrate significant potential in refining inversion accuracy and subsurface characterization (Hancock and Khoshgoftaar 2020; Aleardi et al. 2021; Liu et al. 2023b; Alam et al. 2025; Arif et al. 2025). These techniques effectively analyze large datasets, revealing subtle geological anomalies often overlooked by conventional methods. ML has been successfully applied in landslide susceptibility mapping, subsurface dissolution studies, hydrogeological investigations, groundwater irrigation sustainability, and soil–rock differentiation (Guevara et al. 2017; Dimech et al. 2022; Dick et al. 2025; Akakuru et al. 2025). Furthermore, empirical relationships between seismic velocity and electrical resistivity have been developed using statistical regression and clustering techniques for joint ERT–SRT parameter inversion (Muñoz et al. 2010; Zeng et al. 2018; Delforge et al. 2021; Akingboye and Bery 2023a; Dick et al. 2024b, a). These advancements enhance subsurface characterization across diverse geological settings.

Given these developments, this review evaluates electrical and seismic refraction methods, emphasizing ML-driven inversion and interpretation models, with a detailed synopsis in Fig. 1. Case studies illustrate practical applications of ERT and SRT in tomographic electrode performance, groundwater studies, and engineering investigations. By synthesizing historical and recent advancements, this review bridges research gaps in ML velocity–resistivity relationships. It aims to (i) review theoretical foundations and methodological progress in ERT and SRT, (ii) assess the role of ML in improving data analysis, (iii) examine empirical resistivity and P-wave velocity (Vp) relationships, and (iv) provide a forward-looking perspective on integrated geophysical methods. By consolidating theoretical insights, methodological advancements, emerging trends, and real-world applications, this review serves as a vital resource for geophysicists, researchers, and industry professionals seeking to enhance subsurface characterization through innovative approaches.

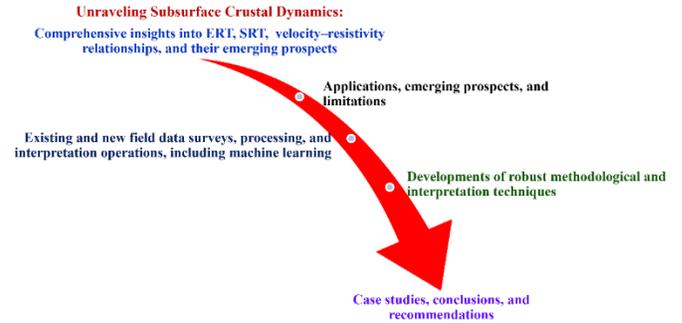

**Fig. 1.** Detailed synopsis of the review, outlining the stages involved in the comprehensive review.

## 2. Insights into electrical methods in earth crustal prospecting

### 2.1 Electrical resistivity method

Electrical methods, including SP, IP, and electrical resistivity techniques, utilize natural or artificially generated fields to characterize surface and subsurface lithologic units (Binley 2015; Akingboye and Ogunyele 2019; Hasan and Shang 2023). Among these, the electrical resistivity method (ERM) employs an artificially generated current to measure the resistivity ($\rho$) or conductivity ($\sigma$) of crustal materials. It uses two current electrodes to inject electrical current into the subsurface, while two potential electrodes measure the resulting voltage (Daily et al. 2005; Abdullah et al. 2022). The measured apparent resistivity ($\rho_a$) is then used to estimate the true resistivity of homogeneous or heterogeneous subsurface materials (Loke et al. 2013; Liu et al. 2020; Akingboye and Bery 2022, 2023a; Alam et al. 2025). Variations in resistivity across lithologies, illustrated in Fig. 2, demonstrate ERM's effectiveness in delineating geological structures. However, assigning precise resistivity values to specific rock types remains challenging due to influencing factors such as density, water content, structural deformities, grain and pore sizes, and weathering intensity. These complexities necessitate careful interpretation of resistivity data.

### 2.1.1 Theories and principles of electrical resistivity method

To ensure efficient performance in subsurface crustal investigations, ERM relies on a fundamental relationship governing electrical resistivity, current, and potential, rooted in Ohm's law (Loke et al. 2013; Binley 2015). This relationship calculates the potential difference for a point current source situated at $X_c$ within a continuous medium, embodying a formulation of Ohm's law that, when combined with the conservation of current, yields Poisson's equation, Eq. 1 (Telford et al. 1990; Loke et al. 2013). Essentially, as illustrated in Fig. 3a, current injected from a single ground electrode on the earth's surface flows radially into the subsurface with uniform resistivity ($\rho$), resulting in a uniform current distribution across hemispherical shells (Loke 2002; Akingboye and Ogunyele 2019).

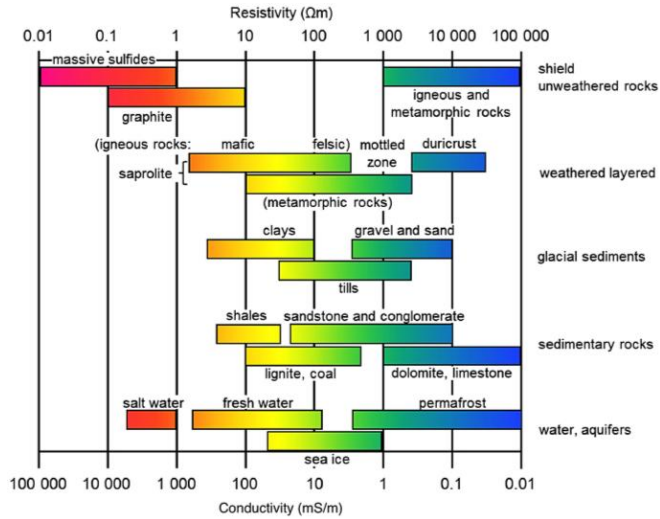

**Fig. 2.** Typical resistivity and conductivity ranges for various geologic materials, including minerals, sediments, and rock types (after Telford et al. 1990). The upper scale represents resistivity (Ωm), while the lower scale represents conductivity (mS/m), both on logarithmic scales. The color gradient indicates variations in resistivity, transitioning from red (low resistivity, high conductivity) to blue (high resistivity, low conductivity).

$$\nabla \cdot \left( \frac{1}{\rho(x,y,z)} \nabla \phi(x,y,z) \right) = -I\delta(X - X_c) \quad (1)$$

Where $\rho(x, y, z)$ is the resistivity distribution, $\phi(x, y, z)$ is the electric potential, $I$ is the injected current, and $\delta(X - X_c)$ represents the point source at the location $X_c$. The Dirac delta function $\delta(X - X_c)$ ensures that the current is applied at a specific point in space.

Solving for $\phi$, the well-known fundamental solution for a single current electrode in a homogeneous half-space, where $r$ is the radial distance from the source, is:

$$\phi(X) = \frac{\rho I}{2\pi r} \quad (2)$$

Theoretically, the relationship governing the four-electrode system (Fig. 3b) in resistivity imaging of surface–subsurface crustal formations follows the same principles as a single-point electrode system. When current is injected into the ground, it spreads radially outward from the injection point, distributing over a hemispherical shell in a homogeneous half-space. At a distance ($r$) from the source, the surface area of this hemispherical shell is given by $A = 2\pi r^2$, allowing for the calculation of the current density ($J$) defined as the total current per unit area (Eq. 3).

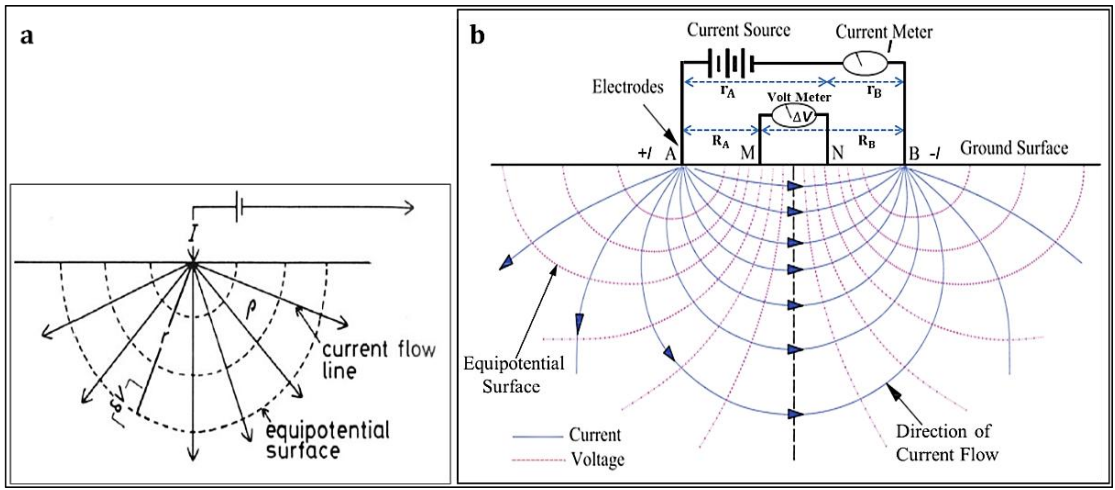

**Fig. 3.** (a) Schematic diagram illustrating a single-point source electrode and the distribution of electric current flow within the subsurface. (b) Typical configuration of resistivity measurements using four electrodes, depicting the direction of the generated current in the earth's subsurface.

$$J = \frac{I}{A} = \frac{I}{2\pi r^2} \quad (3)$$

The charge density can also be given as $J = \sigma E = \frac{E}{\rho}$;

Where $E$ is the electric field (or the gradient of a scalar potential) and is given as:

$$E = -\nabla \phi \quad (4)$$

Hence, relating $J$ with the potential gradient gives:

$$\nabla \phi = -\rho J = -\frac{\rho I}{2\pi r^2} \hat{r} \quad (5)$$

Integrating Eq. 5 gives the potential, as shown below:

$$\phi = \int \nabla \phi \cdot dr = -\int \frac{\rho I}{2\pi r^2} dr = \frac{\rho I}{2\pi r} \quad (6)$$

Equation 6 represents the fundamental solution in a homogeneous medium before generalizing to a four-electrode system. The general equation for the potential difference ($\Delta V$) using a four-electrode measurement system (Fig. 3b) is based on relating the potential $V_M$ at an internal electrode M, which is the sum of the potential contributions of $V_A$ and $V_B$ from the current source at A and sink at B, and that of electrode N (Kearey et al. 2002; Akingboye and Ogunyele 2019). Therefore, the equation is given as:

$$V_M = V_A + V_B \tag{7}$$

From Eq. 6:

$$V_M = \frac{\rho I}{2\pi}\left(\frac{1}{r_A} - \frac{1}{r_B}\right) \tag{8a}$$

Similarly, $V_N = \frac{\rho I}{2\pi}\left(\frac{1}{R_A} - \frac{1}{R_B}\right) \tag{8b}$

Thus, the $\Delta V$ between electrodes M and N is given as:

$$\Delta V = V_M - V_N = \frac{\rho I}{2\pi}\left(\left(\frac{1}{r_A} - \frac{1}{r_B}\right) - \left(\frac{1}{R_A} - \frac{1}{R_B}\right)\right) \tag{9}$$

$$\rho = \frac{2\pi \Delta V}{I\left(\left(\frac{1}{r_A} - \frac{1}{r_B}\right) - \left(\frac{1}{R_A} - \frac{1}{R_B}\right)\right)} \tag{10}$$

$k$ is the geometric factor for any used array, and is mathematically expressed as:

$$k = \frac{2\pi}{\left(\frac{1}{r_A} - \frac{1}{r_B}\right) - \left(\frac{1}{R_A} - \frac{1}{R_B}\right)} \tag{11}$$

For computation purposes, based on Fig. 3b, $k$ can be rewritten as:

$$k = \frac{2\pi}{\left(\frac{1}{r_1} - \frac{1}{r_2}\right) - \left(\frac{1}{r_3} - \frac{1}{r_4}\right)} \tag{12}$$

Note that; $r_1 = r_A$; $r_2 = r_B$; $r_3 = R_A$; and $r_4 = R_B$

Since resistance, R ($\Omega$), is $R = \frac{\Delta V}{I} \tag{13}$

Hence, the apparent resistivity ($\rho_a$) for heterogeneous ground conditions can be mathematically expressed as:

$$\rho_a = kR = k\frac{\Delta V}{I} \tag{14}$$

### 2.2. Electrode array performance evaluation in electrical resistivity surveys

Electrode array configurations in ERM influence data resolution, depth penetration, and sensitivity to geological structures (Abdullah et al. 2018; Akingboye and Bery 2021a). A standard setup comprises four electrodes—two current and two potential electrodes—arranged systematically for land (ground and crosshole) and marine surveys. The choice of array configuration affects noise reduction and lateral/vertical resolution. Electrode arrays also support 3-D/4-D resistivity imaging at various scales (Loke et al. 2022; Oyeyemi et al. 2022; Zakaria et al. 2022). Key considerations for optimizing electrode performance include sensitivity, resolution, electrode spacing, spread length, injected current, probing depth, and noise attenuation (Binley et al. 2015; Akingboye and Bery 2021b; Tomaškovičová and Ingeman-Nielsen 2025). Smaller spacing improves resolution and shallow penetration, while electrode material choice impacts signal quality. For instance, galvanized iron and aluminum electrodes are highly sensitive to noise, whereas copper electrodes, despite their higher cost, ensure efficient current injections due to superior conductivity. Stainless steel electrodes, commonly used in profiling and depth-sounding surveys, offer durability and affordability but are prone to corrosion and lower conductivity (Daily et al. 2005; LaBrecque and Daily 2008). Noise from cultural or self-potential sources can be mitigated using alternating power (Binley and Kemna 2005).

In practice, both conventional and optimized electrode arrays employ different electrode configurations to generate 1-, 2-, 3-, or 4-D resistivity models (Loke 2002; Abdullah et al. 2022; Loke et al. 2022). These configurations influence survey depth and lateral coverage, with certain arrays achieving higher sensitivity to specific structures. For instance, as shown in Fig. 4, the Wenner array, available in alpha (α), beta (β), and gamma (γ) configurations (Fig. 4a–c), is effective in noisy environments and excels in vertical structure delineation, though it has limited horizontal sensitivity (Loke et al. 2013; Merritt 2014). The Schlumberger array (Fig. 4d), similar to the Wenner-α configuration, has closely spaced potential electrodes and is widely used for 1-D resistivity soundings. The Wenner–Schlumberger array (Fig. 4e), a hybrid configuration, provides moderate sensitivity to both horizontal and vertical structures, making it suitable for high-resolution soil-rock models (Loke et al. 2013; Akingboye and Ogunyele 2019).

The dipole-dipole array, mainly in two configurations—normal dipole-dipole (Fig. 4f) and the equatorial dipole-dipole (Fig. 4g)—often employed in ERM and IP surveys, minimizes electromagnetic coupling effects and enhances depth sounding capabilities (Loke 2002; Akingboye 2022). The array is highly sensitive to lateral resistivity variations but less effective in detecting horizontal structures (Rucker et al. 2021b). Other alternatives, such as pole-dipole and pole-pole arrays, offer varying advantages and limitations based on electrode positioning. The pole-dipole array (Fig. 4h), which positions one current electrode at a distance exceeding 20 times the electrode spacing, reduces telluric noise but may produce asymmetric resistivity anomalies (Bery 2016). The pole-pole array (Fig. 4i), characterized by broad lateral coverage and deep probing capability, is more susceptible to telluric noise and may introduce structural artifacts in inversion models, particularly when electrode spacing exceeds 10 m (Zhou and Greenhalgh 2000; Bing and Greenhalgh 2001). Loke and Barker (1996a) optimized the pole-pole array with a cross-diagonal (along x- and y-directions) surveying technique to reduce time and effort while ensuring quality data in 3-D surveys.

The multi-electrode gradient array (Fig. 4j) enhances survey efficiency by maximizing potential readings for a given current injection. The injected current separation $(s + 2)a$ determines the resulting voltage measurements, with the number of potential readings depending on (*s*) (Binley 2015; Zhou et al. 2020). Additionally, the azimuthal square array, designed for directional resistivity analysis, features multiple configurations (α, β, and γ) as shown in Fig. 4k–m. The symmetrical expansion of the square array about its center is depicted in Fig. 4n, while Fig. 4o illustrates its rotation in 15° increments about its midpoint, with the array center serving as the measurement location (Loke et al. 2015c; Martorana et al. 2017; Akingboye 2024b). The depth of investigation for the square array is estimated as 0.451 times the electrode separation distance. To assess resistivity variations accurately, the array is rotated in 15° increments from true north to 180° (Udosen and George 2018).

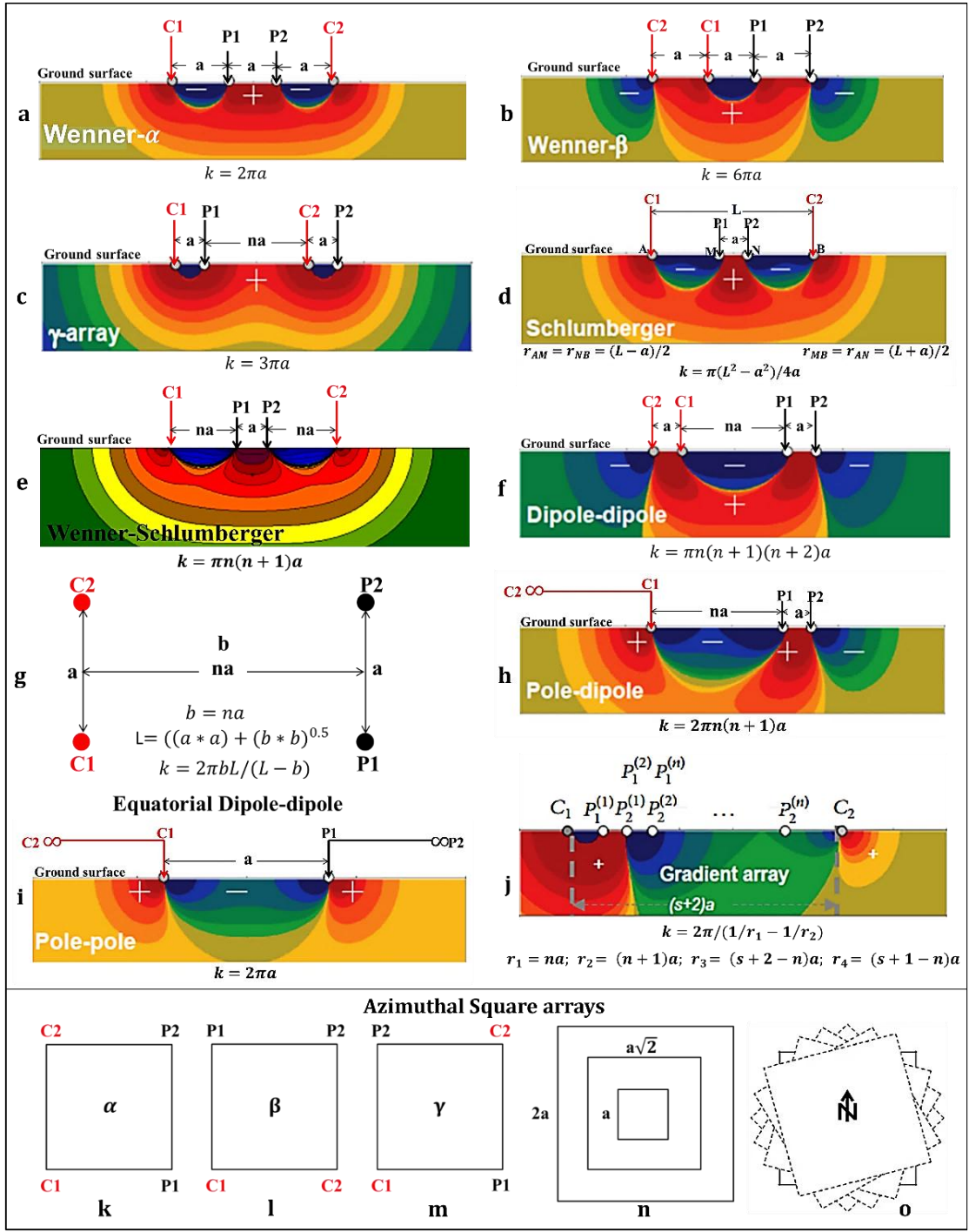

**Fig. 4.** Schematic diagrams of commonly used arrays in ERM with their configurations and sensitivity patterns alongside their computed geometric factors. The current electrodes are denoted as C1 and C2, while P1 and P2 represent the potential electrodes. $n$ signifies a positive integer value for the dipole separation factor, $a$ denotes the electrode spacing, $k$ represents the geometric factor, and ∞ indicates a larger electrode separation approximately 20 times the normal spacing (adapted from Loke 2002; Akingboye and Ogunyele 2019).

*2.3 One-dimensional electrical resistivity sounding and modeling.*

The effectiveness of the 1-D electrical resistivity sounding method, particularly in subsurface investigations, remains a subject of debate due to its inherent limitations. Although the Schlumberger array is commonly employed due to its efficiency in detecting vertical resistivity variations, the Wenner array and, less frequently, the dipole-dipole array have also been applied in specific contexts (Singh 2005; Akintorinwa et al. 2020; Akingboye 2022). However, the choice of an array is critical, as it influences resolution, depth penetration, and sensitivity to noise, factors that are often overlooked in standard survey designs (Akingboye et al. 2023). A major challenge in 1-D resistivity processing lies in the reliability of $\rho_a$ values, which are plotted against AB/2 or station distance on bi-logarithmic or log-log graph sheets for quality assurance. While such graphical methods help in identifying anomalies, their reliance on visual interpretation introduces subjectivity, leading to potential errors (Oladapo et al. 2004). Despite advances in inversion techniques, iterative curve-matching approaches used in software-based modeling still depend heavily on initial parameter selection, which can influence the final resistivity estimates. This raises concerns regarding the reproducibility and accuracy of VES-derived subsurface models, particularly in heterogeneous geological settings.

Typical examples of 1-D models used for interpreting subsurface scenarios are depicted in Fig. 5. The classification of VES curves into A-, Q-, H-, and K-types (Fig. 6) provides a fundamental framework for resistivity interpretation. However, these curve types often oversimplify complex subsurface conditions, leading to potential misinterpretations, especially when lateral heterogeneities are present. While the method has been widely applied in hydrogeological and geotechnical studies (Oladapo et al. 2004), its single-point probing nature limits its ability to resolve lateral variations in stratigraphy. This limitation becomes particularly problematic in regions with sharp lithological contrasts or complex structural formations, where 1-D models may produce misleading subsurface characterizations. A more robust approach would involve integrating multiple VES soundings along profiles for improved layer correlation, or ideally, combining 1-D results with 2-D/3-D resistivity imaging techniques to enhance spatial resolution.

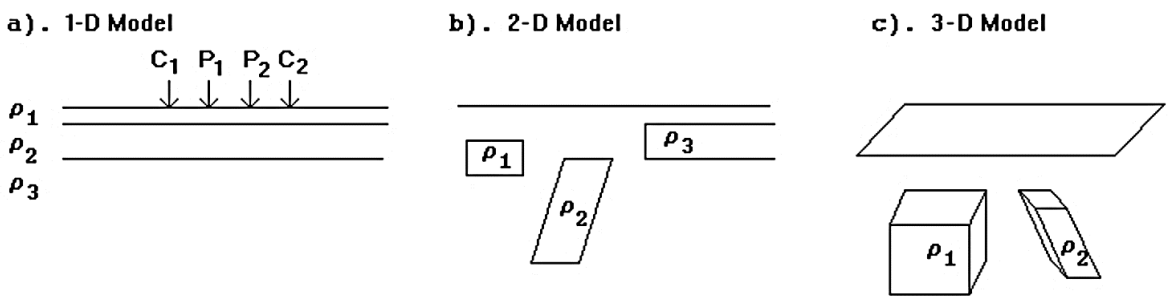

**Fig. 5.** Typical models used in the interpretation of resistivity measurements (after Loke 2002).

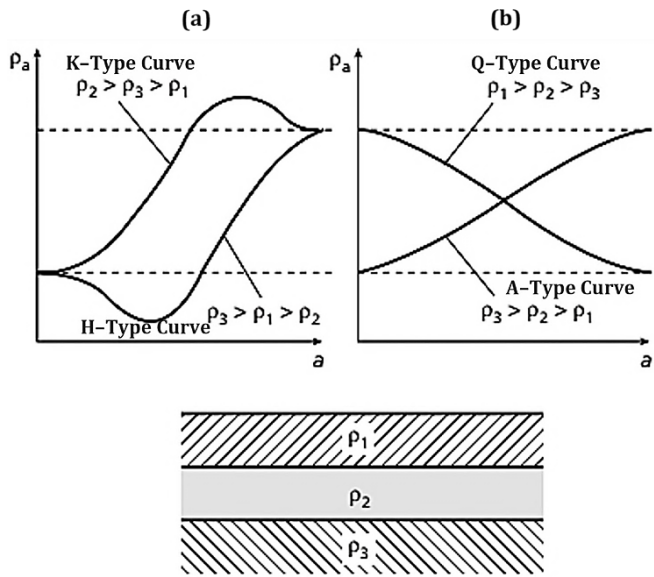

**Fig. 6.** Schematic model of the main VES curve types observable over a three-layer horizontal body due to the variation of apparent resistivity with electrode separation (modified after Kearey et al. 2002).

*2.4 Induced polarization method*

The IP method, measured in msec (or $mV/V$), developed by Conrad Schlumberger in the 1920s (Schlumberger 1920), remains a crucial geophysical technique despite challenges such as electrode polarization, data noise, and interpretation ambiguity. It measures the delayed voltage response after an applied current ceases, capturing chargeability, which depends on grain size, mineral composition, and electrolyte properties in pore spaces (Dahlin et al. 2002; Binley 2015; Biosca et al. 2020). However, distinguishing materials like sulfides, clays, and graphite requires complementary data (Dahlin et al. 2002; Revil et al. 2022), as chargeability responses vary with subsurface conditions (Slater et al. 2010; Piegari et al. 2023). Compared to ERM, IP is more sensitive to fine-scale variations at the fluid-grain interface, resolving ambiguities in low-resistivity contrasts (Kemna et al. 2004; Rucker et al. 2021b). While integrating IP and ERM enhances subsurface characterization (Slater & Glaser 2003), practical constraints such as increased acquisition time, power requirements, and telluric noise must be considered (Kearey et al. 2002). IP is widely applied in mineral exploration, hydrogeological mapping, and engineering site investigations, including saltwater intrusion studies and pollutant detection (Martínez et al. 2019; Rey et al. 2020; Kumar et al. 2021).

Survey design significantly impacts IP data reliability. Commonly employed electrode configurations—dipole-dipole, pole-dipole, and Schlumberger arrays—vary in depth penetration, resolution, and noise sensitivity (Binley and Kemna 2005; Revil et al. 2022). Non-polarizing electrodes like copper-copper sulfate minimize polarization effects, while stainless steel or copper electrodes balance practicality and accuracy (Akingboye and Bery 2021a). Electrode spacing ranges from 0.5 m in engineering studies to 300 m in reconnaissance surveys (Kearey et al. 2002; Rucker et al. 2021a). External noise sources, including telluric currents and barren rock effects, necessitate careful data acquisition strategies (Daily et al. 2005). IP measurements utilize either time-domain or frequency-domain methods. Time-domain IP records voltage decay after current cessation, as shown in Fig. 7, reflecting electrical charge polarization within pore fluids and grain boundaries (Binley and Kemna 2005). Frequency-domain IP measures phase-shifted voltage and employs the percent frequency effect (PFE) as a key parameter (Günther and Martin 2016). Injected currents in IP are typically higher than in ERM, particularly when using dipole-dipole arrays. However, IP shares ERM's limitations, such as sensitivity to water-filled shear zones and graphite-rich sediments, which may lead to misinterpretations with economic implications (Li and Oldenburg 2000; Slater et al. 2010; Lenhare and Moreira 2020).

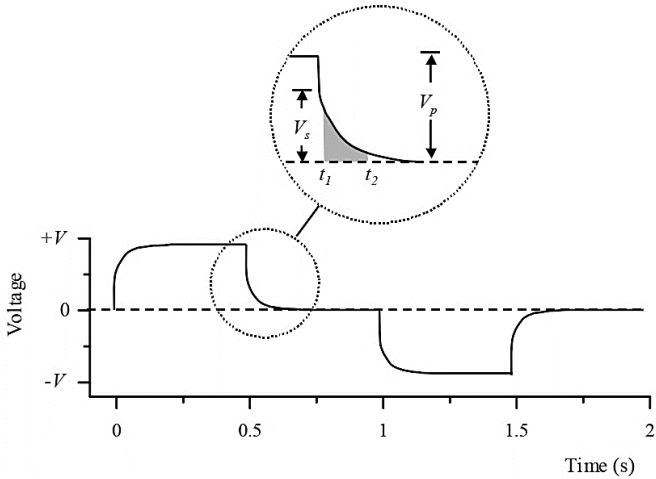

**Fig. 7.** Typical measurement of time-domain induced polarization (after Binley and Kemna 2005).

As defined by Seigel (1959), the mathematical theory supporting the apparent chargeability ($m_a$) measurement of IP is expressed in Eq. 15.

$$m_a = \frac{V_p}{V_s} \qquad (15)$$

The primary voltage is denoted as $V_p$ while $V_s$ is the secondary voltage measured in volt. It is important to note that measuring the $V_s$ in the field is quite difficult; hence an integral measure of $m_a$, as illustrated in Fig. 7 and mathematically represented in Eq. 16, is used. The chosen time window ($t_1$ to $t_2$) employed will affect the measured $m_a$, as the PFE impacts the selected injection frequencies and impedance phase angle.

$$m_a = \frac{1}{t_2 - t_1} \frac{1}{V_p} \int_{t_1}^{t_2} V(t) dt \qquad (16)$$

IP measurement modes commonly made are the PFE and the metal factor (MF) (Kearey et al. 2002). PFE and MF are mathematically defined as:

$$PFE = \frac{(\rho_{0.1} - \rho_{10})}{\rho_{10}} \qquad (17)$$

$$MF = 2\pi \times 10^5 \frac{(\rho_{0.1} - \rho_{10})}{\rho_{0.1} \rho_{10}} \qquad (18)$$

Where $\rho_{0.1}$ and $\rho_{10}$ are the apparent resistivities at measuring frequencies of 0.1 and 10 Hz.

All the itemized parameters provide quantitative frameworks for analysis. However, traditional approaches often rely on simplified geometries (e.g., spheres, dykes, or horizontal layers), which may not accurately represent complex subsurface structures (Revil et al. 2022). Laboratory modeling and simulations offer alternative interpretations of IP anomalies (Akingboye and Bery 2021a). Advancements in IP interpretation integrate computational techniques, including ML-based models (Piegari et al. 2023; Bala et al. 2024) and 3-D inversion (Rucker et al. 2021b; Martorana et al. 2023). Bala et al. (2024) recently demonstrated the potential of resistivity–chargeability relationships in sedimentary–basement terrains using a supervised statistical approach in Langkawi, Malaysia. Such innovations enhance IP interpretation, particularly in environmental and geotechnical studies. Despite progress, standardizing methodologies and ensuring automated models' adaptability across geological settings remain challenging. While IP is valuable, integrating hybrid ML and complementary geophysical methods can enhance accuracy. However, field constraints, data complexities, and chargeability variability require careful application and rigorous quality control.

## 2.5 Self-potential method

Though this study does not delve into SP, it is crucial to briefly touch upon its arrays and procedures for clarity in distinguishing among electrical survey arrays. Unlike IP and ERM, SP employs a distinct methodology. SP, based on measuring natural potential differences due to subsurface electrochemical reactions, utilizes specialized non-polarizable electrodes connected to a high-impedance millivoltmeter (Kearey et al. 2002; Jackson 2015). These electrodes, termed "porous pot" electrodes, contain either copper-copper sulfate (Cu–CuSO$_4$) or silver-silver chloride (Ag–AgCl) solution, the former for land surveys and the latter for marine/water surveys (Telford et al. 1990; Hasan and Shang 2023). SP involves passive measurement of electric potential at the surface or in boreholes, offering a cost-effective and rapid alternative with simple equipment (Mao et al. 2015). While useful for detecting changes in groundwater flow, chemistry, and temperature, its interpretation remains complex, and its limited depth penetration (≈30 m) restricts its role in exploration compared to other electrical methods (Jackson 2015; Mao et al. 2015).

## 3. Electrical resistivity tomography

ERT has evolved into a powerful tool for subsurface imaging, providing detailed insights into geological structures across various environments, including land, boreholes, and underwater settings (Daily et al. 2005; Perri et al. 2020). Advances in survey techniques, inversion algorithms, and multi-physics integration with seismic or borehole data have significantly improved its accuracy and efficiency (Wilkinson et al. 2014; Abdullah et al. 2019, 2022). Moreover, the incorporation of ML, AI, and statistical modeling enhances data interpretation, refining resistivity imaging for complex subsurface conditions (Aleardi et al. 2021; Liu et al. 2023b, a). A key advancement, time-lapse ERT (TLERT), enables 3- and 4-D monitoring of dynamic subsurface changes, making it invaluable for environmental and geotechnical applications (Palis et al. 2017; Loke et al. 2022; Fang et al. 2023). Operating on the same principles as electrical resistivity methods, ERT employs four electrodes to inject current and measure potential differences, with larger electrode spacing allowing deeper penetration (Loke et al. 2013; Binley 2015; Doyoro et al. 2025). The resulting apparent resistivity measurements generate pseudosections and inverse resistivity models, effectively delineating soil and rock conditions for geological assessment (Dahlin and Loke 2018; Szalai et al. 2022; Raji et al. 2023).

Compared to conventional ERM, ERT offers several advantages, including faster data acquisition, reduced operational costs, and minimized manual electrode switching (Loke et al. 2014a; Delgado-Gonzalez et al. 2023). Its high-resolution imaging and cost-effectiveness make it suitable for both small- and large-scale investigations (Dimech et al. 2022). However, ERT interpretations are influenced by factors such as lithological variations, pore fluid chemistry, and soil moisture content (Daily et al. 2005; Loke et al. 2013). Imaging resolution is also constrained by electrode spacing, subsurface heterogeneity, and inherent ambiguities in inversion models (Daily et al. 2005; Binley 2015). Integrating ERT with IP and SP methods can help mitigate these challenges (Kumar et al. 2020; Hasan and Shang 2023), while closer electrode spacing and supplementary geotechnical data further improve inversion constraints (Wilkinson et al. 2006; Szalai et al. 2020; Akingboye et al. 2024). Borehole logs remain essential for resolving ambiguities and distinguishing true geological structures from inversion artifacts (Dick et al. 2024a).

ERT is widely used in subsurface investigations across multiple disciplines. In geotechnical studies, it is instrumental in mapping underground structures and monitoring groundwater flow related to slope failures and landslides (Palis et al. 2017; Bièvre et al. 2021; Zakaria et al. 2022). Environmental applications include delineating contaminant plumes, assessing groundwater movement, and monitoring subsurface pollution (Mao et al. 2015; Akingboye et al. 2022; Hasan and Shang 2023). In agrogeophysics, it aids in evaluating soil properties, mapping water-nutrient exchange zones, and studying plant root distributions (Cheng et al. 2019; Cordero-Vázquez et al. 2021). Archaeologists use ERT to detect buried structures and historical remains (Nero et al. 2016; Raji et al. 2023), while in mineral exploration, it assists in delineating subsurface ore bodies (Martínez et al. 2019; Sendrós et al. 2020).

*3.1 ERT: field survey design and data acquisition techniques*

ERT surveys can be conducted in 2-, 3-, and 4-D configurations, allowing for versatile applications on land, in crosshole, or underwater (marine ERT), depending on the specific parameters of interest in the subsurface (Merritt et al. 2014; Abdullah et al. 2018; Loke et al. 2022; Akingboye and Bery 2023a). Unlike other survey types, 4-D ERT is distinguished by its ability to monitor resistivity changes over time, making it particularly useful for tracking dynamic subsurface processes (Merritt 2014; Kuras et al. 2016; Loke et al. 2022). The equipment and methodologies for ERT surveys are similar to those used in ERM, utilizing multi-electrode arrays and advanced data loggers (Fig. 8a) to facilitate simultaneous data acquisition for both conventional and modified arrays, as illustrated in Fig. 4 (Crook et al. 2008; Abdullah et al. 2018, 2022). Modern instruments, such as the ABEM Terrameter LS, integrate the Terrameter and Switching Unit into a single device, streamlining measurements and enhancing electrode automation.

2-D ERT land surveys are typically conducted using two or four sets of multicore cable reels (Fig. 8a–b), with evenly spaced take-outs connected to grounded electrodes. The electrodes not only inject currents and measure potential differences but also counteract electrode polarization, which can introduce noise during measurements (Daily et al. 2005; Zhou 2019). The initial profile length in measurement surveys is determined by the length of multicore cables. If the survey area exceeds this length, a roll-along technique can be employed, as shown in Fig. 8c, allowing for continuous data collection in both directions. The electrical measurements sequence (Fig. 8d), array type, and current injection levels are controlled by the resistivity meter, with a switching unit managing the injection and measurement electrodes. The resistivity data are gathered at various depths by measuring resistivity at multiples of the electrode spacing ($a$), typically up to 16 times or more (referred to as *n levels*) (Dahlin 1996; Loke 2002; Loke et al. 2013).

In 3-D surveys, pole-pole, pole-dipole, and dipole-dipole arrays are commonly used for their superior resolution at survey grid edges. However, optimized arrays have further improved data quality and efficiency (Tejero-Andrade et al. 2015; Abdullah et al. 2022). Field layouts typically adopt a square or rectangular grid pattern with uniform electrode spacing along both x and y directions (Fig. 8c), tailored to the geometry of the target body (Loke 2002). Based on reports of Loke (2002), 3-D resistivity measurements employ various methods: (1) collecting data along potential directions using a rectangular grid; (2) measuring along all grid lines, sometimes with limited angular coverage; (3) using a multi-electrode system to collect data along two directions with restricted nodes; and (4) acquiring data along parallel 2-D survey lines, later merged through inversion. Methods (1) and (2) generally provide better angular coverage, whereas (3) and (4) may lack critical directional information. Meanwhile, 4-D ERT, or TLERT, has become essential for monitoring complex terrains and tracking spatiotemporal variations in moisture or subsurface properties over extended periods

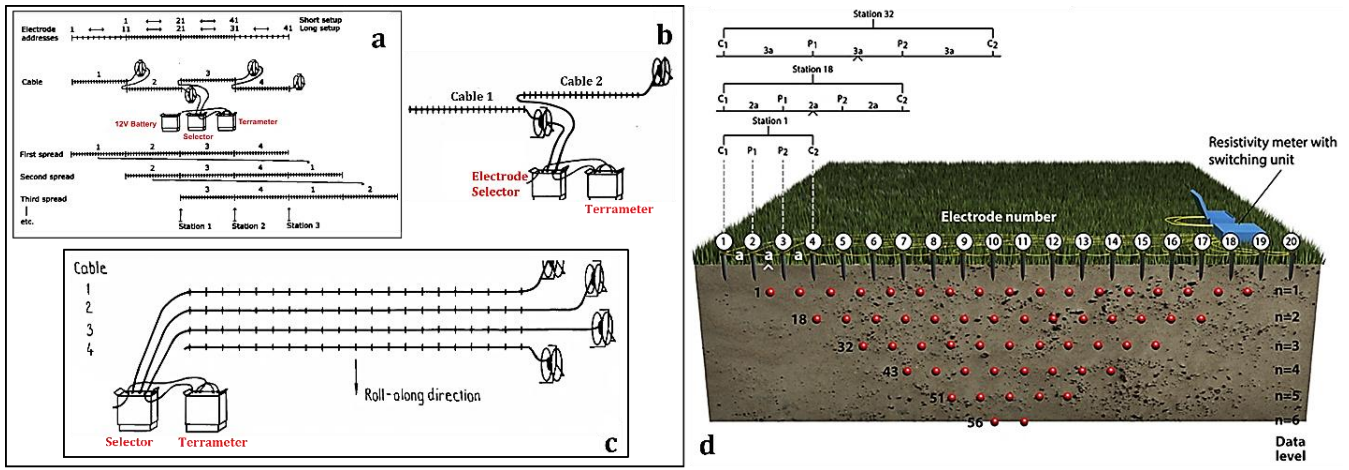

**Fig. 8.** Schematic diagrams for a 2-D typical system layout (a) using four and (b) two multicore cable reels. (c) A typical 3-D ERT survey layout showing a roll-along direction. (d) Schematic image of multi-electrode measurement sequence for a 2-D survey (after Loke et al. 2013).

(Bièvre et al. 2021; Loke et al. 2022).

## 3.2 Optimized electrode arrays in ERT surveys

Recent advancements in automated procedures have greatly improved electrode configuration optimization for 3-D surveys, leading to highly efficient arrays such as the L and Corner, Compare R, and noise-weighted perimeter arrays (Wilkinson et al. 2006; Loke et al. 2015a; Abdullah et al. 2022). Despite limited comparative studies, these optimized arrays enhance model resolution and mitigate computational challenges. Wilkinson et al. (2006) introduced the Compare R array to optimize 2-D resistivity surveys, later validated by Loke et al. (2010b) through model resolution comparisons. Further improvements include parallel programming for faster processing (Loke et al. 2010a) and electrode polarization noise reduction (Wilkinson et al. 2012).

Ongoing research continues to refine these techniques, with Abdullah et al. (2018) assessing modified Compare R arrays. Optimized arrays are particularly valuable in complex urban environments, improving resolution for surficial and subsurface structures. The L array, introduced by Baker et al. (2001), arranges electrodes along orthogonal lines, while the "horse-shoe geometry" (Argote-Espino et al. 2013; Chávez et al. 2014) integrates L, equatorial-dipole, and minimum-coupling (MC) arrays for archaeological assessments. Studies have applied L and Corner array along engineering structure perimeters to map subsurface defects (Chavez-Hernandez et al. 2011; Tejero-Andrade et al. 2015). However, these methods perform best for sharp-edged perimeters, such as rectangles, and are less suited for circular boundaries. To address this, Loke et al. (2015c) modified the Compare R (Loke et al. 2014b) to optimize electrode arrays for confined-area perimeter surveys.

For example, Abdullah et al. (2019a) conducted a study at Universiti Sains Malaysia (USM) to enhance 3-D resistivity survey resolution in confined areas using optimized perimeter arrays developed by Loke et al. (2015b). They introduced a noise-weighted optimized perimeter array to reduce noise and improve imaging quality. Figure 9a–b illustrates possible 2-D perimeter survey arrays forming closed loops, generating a sequence of optimized perimeter arrays (Fig. 9c–f). To assess performance, synthetic models and field datasets were used to compare the standard L and Corner array (Tejero-Andrade et al. 2015) with the new perimeter arrays. A key challenge in perimeter surveys is resolution degradation at greater depths due to increased electrode spacing. To address this, the study applied Loke et al. (2015b) weighting procedures, emphasizing cells within the survey loop to enhance model resolution. Field models generated using the L and Corner array, optimized perimeter arrays, and noise-weighted optimized perimeter arrays (Fig. 10a–c) revealed similar lithological conditions at different depths. However, the optimized and noise-weighted arrays provided superior resolution of deeper anomalies. This advancement significantly improves 3-D surface–subsurface characterizations for engineering and environmental geophysics.

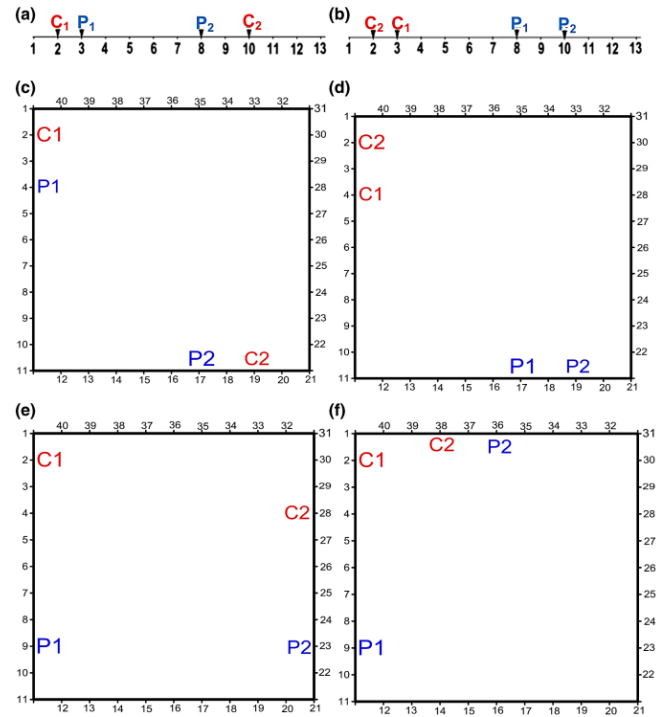

**Fig. 9.** Typical array configuration of (a) alpha and (b) beta types for a 2-D survey traverse. (c) alpha and (d) beta arrays with electrodes on perpendicular segments of a square loop; (e) alpha array with electrodes on opposing segments; (f) alpha array with electrodes on the first and fourth segments (adapted from Abdullah et al. 2019a).

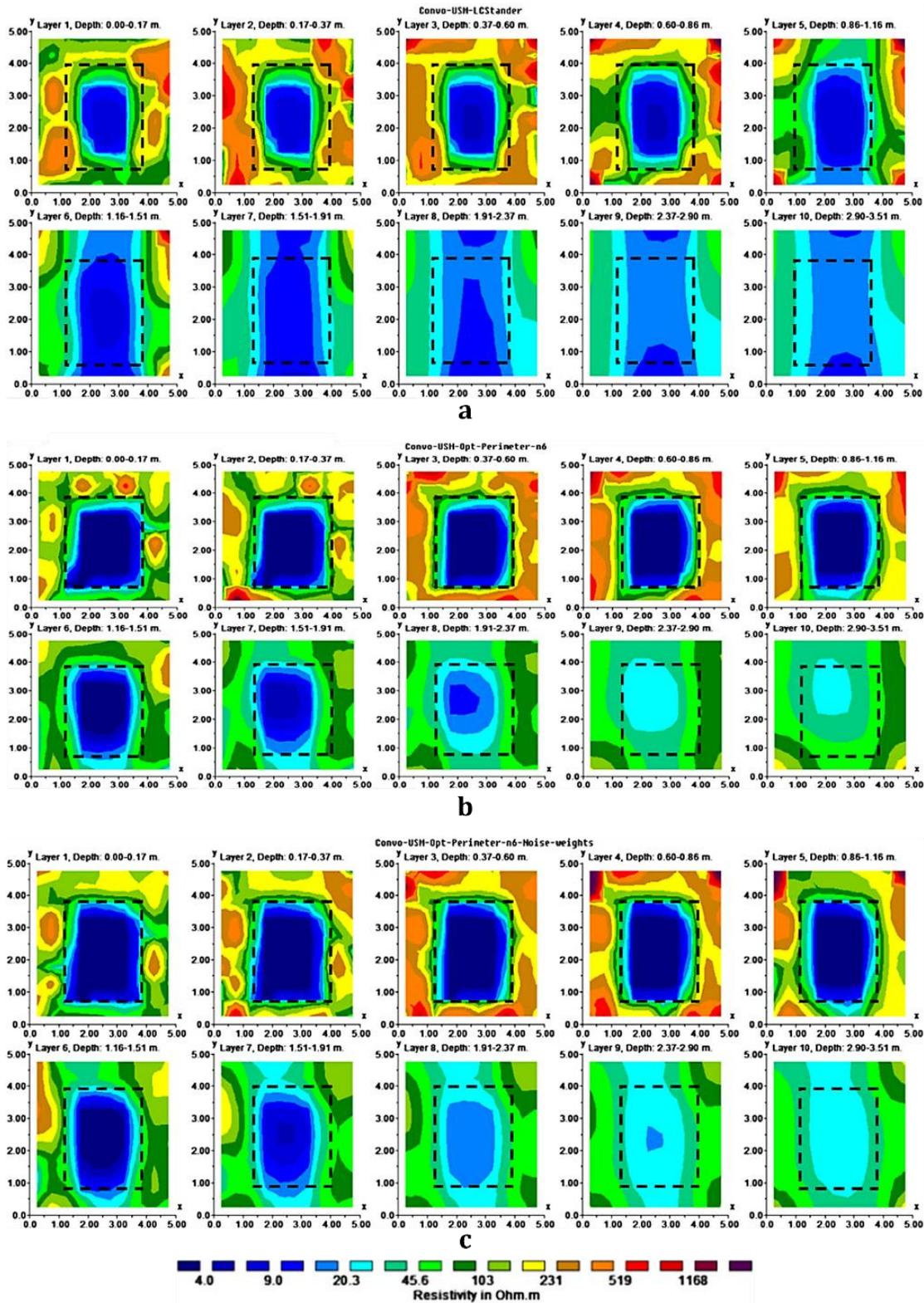

**Fig. 10.** Resistivity models for (a) standard "L and Corner", (b) optimized perimeter, and (c) the noise-weighted optimized perimeter arrays, for real-time field survey within the Universiti Sains Malaysia (adapted from Abdullah et al. 2019a).

## 3.3 Crosshole (borehole-based) electrical surveys

Crosshole electrical surveys address the limitation of surface electrical imaging by enhancing subsurface imaging resolution through various quadrupole combinations of current and potential electrodes placed within boreholes or between boreholes and the surface (Slater et al. 2000; Kemna et al. 2004; Yang et al. 2021), Fig. 11a. Crosshole electrical imaging can be employed for ERM and IP, particularly ERT. To achieve optimal resolution, the separation between boreholes should not exceed approximately 0.75 times the length of the borehole array (Daily et al. 2005). Borehole layouts, whether regular or irregular, are chosen based on the specific site conditions and investigative requirements. However, accurate positioning of borehole electrodes presents challenges, necessitating consideration of random and systematic electrode offsets and borehole deviations (Slater et al. 2000; Wilkinson et al. 2006).

Recent advancements integrate ERT data with IP measurements or borehole logs to delineate lithological boundaries, particularly distinguishing clay from water-saturated soils (Slater and Lesmes 2002). These methods effectively characterize pore fluid chemistry, void spaces, soil water content, and other subsurface properties (Binley and Kemna 2005). Further improvements involve combining advanced technologies, methodologies, and algorithms to enhance crosshole ERT for subsurface imaging and characterization (Doetsch et al. 2012). These developments include integrating fiber-optic distributed acoustic sensing (DAS) and electromagnetic techniques with crosshole ERT to optimize electrode positioning and enable real-time monitoring of subsurface conditions (Bergmann et al. 2016).

Ongoing efforts focus on combining crosshole ERT with complementary geophysical and geotechnical methods, such as seismic tomography, ground-penetrating radar (GPR), and cone penetration testing, to improve subsurface characterization. This multi-method approach enhances data interpretation, providing greater accuracy in assessing complex geological formations, hydrological processes, and environmental conditions (Li and Oldenburg 2000; Slater and Lesmes 2002; Dimech et al. 2022). Crosshole ERT is particularly valuable for TLERT, facilitating the monitoring of groundwater or moisture fluctuations and tracer concentrations (Dimech et al. 2022). A representative survey model illustrating surface and crosshole electrode configurations for tracer monitoring is shown in Fig. 11b. To improve the surface response of anomalous bodies, Yang et al. (2021) proposed a 3-D inversion technique incorporating damped least-squares, finite difference simulations, and the incomplete Cholesky conjugate gradient method.

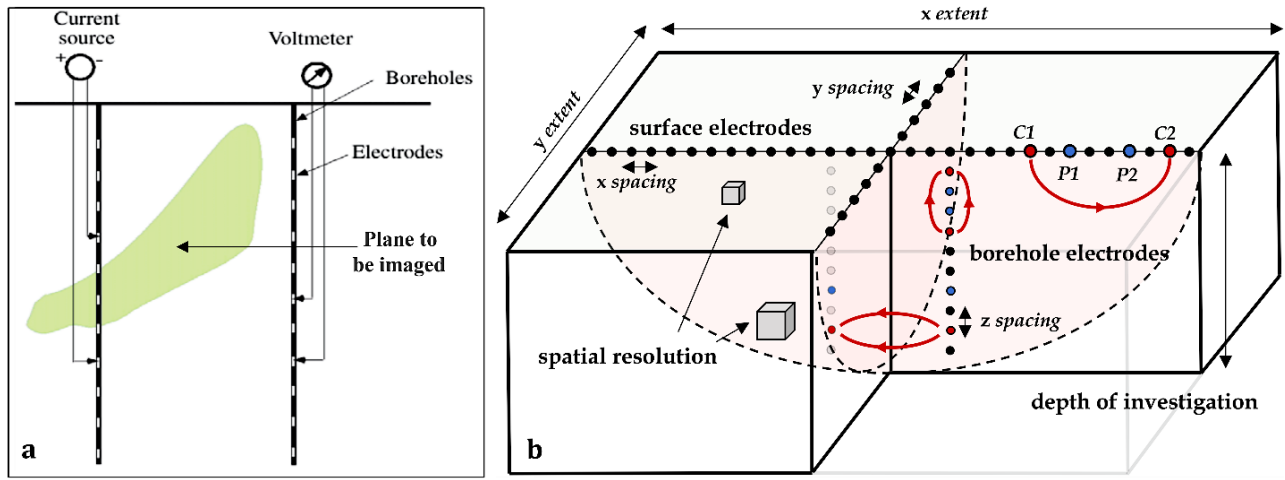

**Fig. 11.** (a) Schematic of crosshole electrical resistivity survey measurement layout (modified after Daily et al. 2005). (b) Schematic crosshole spatiotemporal parameters of TLERT survey approach (adapted from Dimech et al. 2022).

## 3.4 Surface- and sea-bottom-towed marine-based ERT surveys

Marine resistivity is an essential noninvasive tool in geotechnical, geoenvironmental, and coastal studies, offering insights into sub-bottom geology, sedimentary structures, and hydrological processes (Snyder et al. 2002; Day-Lewis et al. 2006; Orlando 2013). It aids in assessing engineering foundation suitability, coastal erosion, habitat mapping, and groundwater discharge (Snyder et al. 2002; Dahlin and Loke 2018; Benevides et al. 2024). Mobile ERT imaging systems have been adapted for both land and underwater surveys (Goto et al. 2008; Passaro 2010; Hermans and Paepen 2020), as shown in Fig. 12a–b, each with distinct advantages and challenges in exploring underwater lithologies.

Continuous resistivity profiling (CRP) remains the standard approach for underwater ERT imaging, utilizing floating electrode arrays towed along the water surface (Snyder et al. 2002; Day-Lewis et al. 2006), as illustrated in Fig. 12a. Dipole-dipole arrays are commonly employed due to their efficient geometry, rapid data acquisition, and compatibility with multichannel resistivity meters (Day-Lewis et al. 2006). Additionally, dipole-dipole and multiple gradient arrays effectively detect fracture zones in resistive bedrock, offering improved depth penetration and sensitivity to vertical resistivity boundaries (Loke et al. 2013). Floating electrodes are most effective when the water column is <25% of the total investigation depth, facilitating deployment and increasing productivity (Akingboye and Ogunyele 2019).

Data interpretation with floating electrodes treats the water layer as either an unknown or fixed resistivity layer, as shown in Fig. 12c,

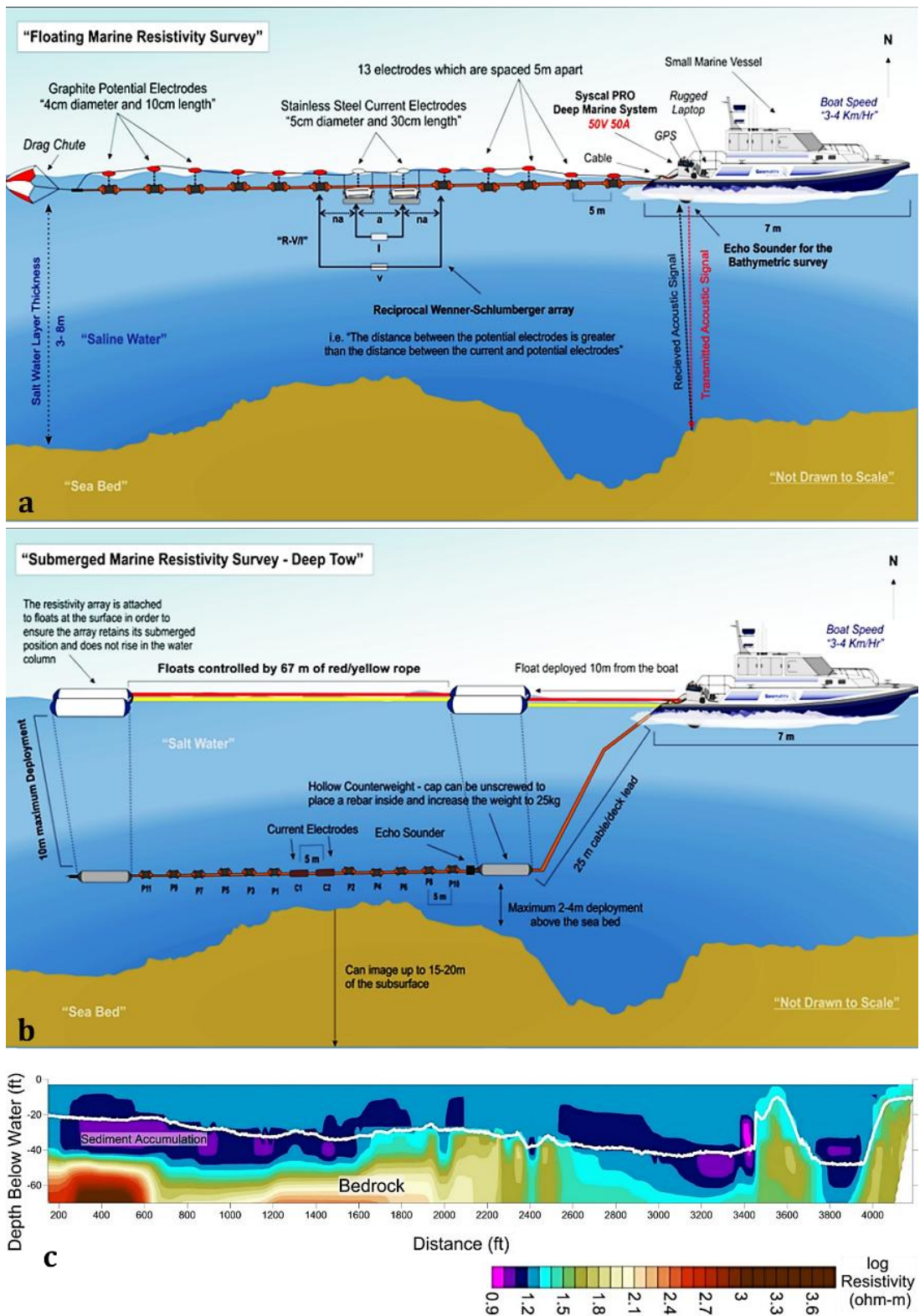

**Fig. 12.** Schematic diagrams of (a) the floating and (b) the submerged mobile underwater ERT survey system (adapted from Geomatrix Earth Science Ltd 2015). (c) A typical resistivity model section of Patagonia Lake, Arizona, derived from a CRP marine survey conducted using a floating-type marine resistivity system (adapted from HGI 2021).

making it suitable for mapping conductivity variations or freshwater-saltwater interfaces (Passaro 2010; Ronczka et al. 2017; Wang et al. 2018). However, in saline environments, resolution decreases as water depth increases, limiting current penetration. To mitigate these challenges, sea-bottom electrode configurations (Fig. 12b) are employed, placing electrodes directly on the waterbed for enhanced resolution and deeper penetration (Loke et al. 2013). This method, however, requires specialized marine cables resistant to mechanical stress and lower survey speeds (Day-Lewis et al. 2006; Orlando 2013). Notably, marine ERT data processing necessitates precise parameter considerations due to weak current distribution, deep target layers, extended inversion times, and resistivity contrasts between water and bedrock (Loke et al. 2013; Orlando 2013).

## 4. Insights into seismic methods in surface–subsurface crustal prospecting

Seismic waves, originating from natural or artificial sources, propagate as elastic strain energy, forming pulses essential to seismic surveys across various frequencies (Côrte et al. 2020; Doyoro et al. 2025). Seismic surveys are categorized into reflection and refraction methods, illustrated in Fig. 13. Reflection surveys analyze travel times of waves reflecting off subsurface interfaces, producing travel time curves (Fig. 13a) that reveal geological structures. In contrast, refraction surveys measure the travel times of refracted waves (Fig. 13a–b), utilizing acoustic impedance contrasts at geological boundaries (Moorkamp et al. 2013; Akingboye and Ogunyele 2019). Reflection methods are widely used for deep crustal studies, including mapping carbon sequestration sites and crustal dynamics, especially in deep waters. Refraction methods, covering shallow to intermediate depths, are vital in hydrogeology, environmental assessments, and geoengineering, often refining reflection interpretations. Advanced data processing techniques eliminate noise and direct waves, enhancing accuracy. Recent innovations in integrated algorithms and three-component seismometers have significantly improved seismic imaging resolution (Li et al. 2020; Yang et al. 2020).

Seismic pulse velocities are governed by the elastic moduli and densities of subsurface materials, classified into body and surface waves. Body waves, comprising compressional (P-waves) and shear waves (S-waves), propagate through solid volumes, with P-waves traveling faster due to their compressional-dilatational motion along the wave direction (Lin et al. 2015; Salleh et al. 2021). S-waves, moving perpendicular to wave travel, induce shear strain. Surface waves, including Rayleigh and Love waves, travel along the solid boundary (Telford et al. 1990; Kearey et al. 2002). Generally, mantle rocks, being denser and more rigid than crustal rocks, allow for faster P- and S-wave propagation (Telford et al. 1990; Earle 2019), as shown in Fig. 14. Seismic-wave velocities increase with depth due to rock compression, while partial melting slows propagation, with complete liquefaction dramatically reducing P-wave velocity and halting S-wave transmission (Earle 2019). This study focuses on seismic refraction velocity, emphasizing recent advancements in SRT for measuring, processing, and interpreting P-wave refraction velocity (Vp).

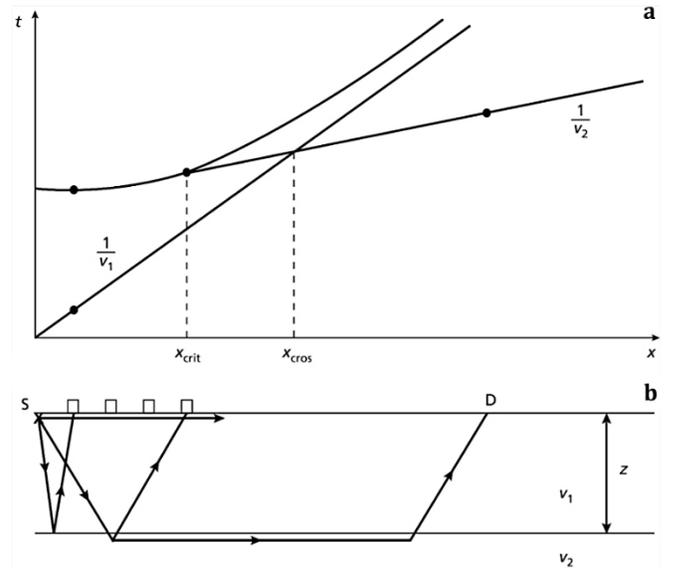

**Fig. 13.** (a) Schematic representation of a basic two-layered travel time curves model. (b) Illustration of a typical seismic acquisition setup depicting direct, reflected, and refracted seismic waves from a source (S) to a surface geophone detector (D) (adapted after Kearey et al., 2002). z, $V_1$, $V_2$, $X_{crit}$, and $X_{cros}$, denote depth of the layer, velocity of the first layer, velocity of the second layer, critical distance, and crossover distance, respectively. t is the intercept time. $1/V_1$ and $1/V_2$ are the gradients of the first and second layers, respectively.

The mode of propagation of a P-wave in isotropic and homogeneous substances is always longitudinal. Therefore, Vp in a homogeneous isotropic medium is given as:

$$Vp = \sqrt{\frac{K+\frac{4}{3}\mu}{d}} = \sqrt{\frac{\lambda+2\mu}{d}} \quad (19)$$

Where $K$ represents the bulk modulus, which is the modulus of incompressibility; $\mu$ is the shear modulus (i.e., the rigidity modulus); $d$ denotes the bulk density of the material through which the wave propagates, and $\lambda$ is Lamé's constant. The bulk density $d$ exhibits the least variation, controlled by parameters $K$ and $\mu$. The P-wave elastic modulus ($Mp$) of the propagating waves can be estimated as:

$$Vp = \frac{Mp}{d} \quad (20)$$

The velocity of the accompanying S-wave (Vs), is given as:

$$Vs = \sqrt{\frac{\mu}{d}} \quad (21)$$

Equations 19 and 21 suggest that Vp is faster in the same medium than Vs. Hence, the ratio of both waves as given in Eq. 22, is referred to as Poisson's ratio ($\varphi$) for the medium, illustrating the elastic (or deformation) properties of the propagated medium.

$$\frac{Vp}{Vs} = \sqrt{\frac{2(1-\varphi)}{1-2\varphi}} \quad (22)$$

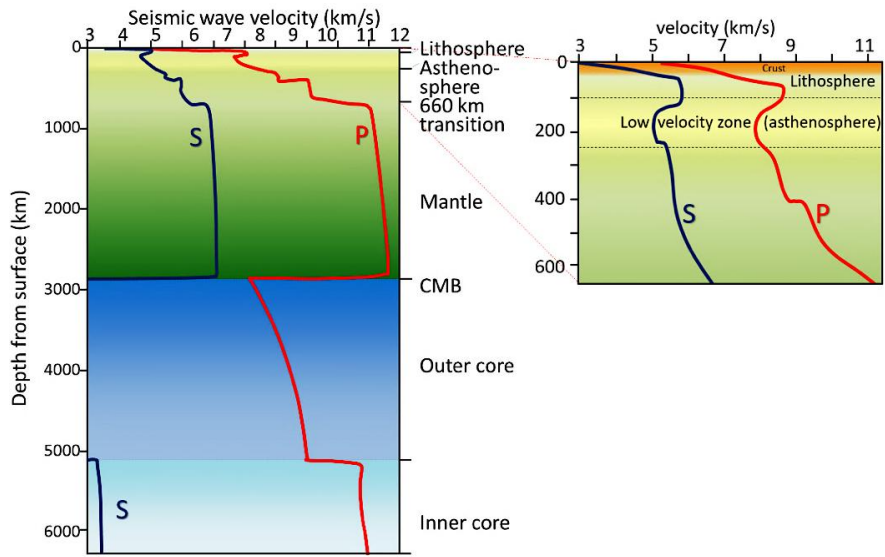

**Fig. 14.** Image of seismic velocity variations of P-wave (red) and S-wave (blue) deep within the earth and the projected magnification image of the upper 660 km depth (adapted from Earle 2019).

*4.1 Seismic refraction tomography*

Recent advancements in SRT, driven by powerful computer-assisted technologies, have significantly improved earthquake location determination and seismic body wave travel-time analysis. Like conventional seismic refraction methods, SRT utilizes seismic energy from external sources (Jianliang et al. 2022; Akingboye 2023; Sujitapan et al. 2024). However, SRT benefits from advanced survey loggers and geophones, such as the ABEM Terraloc Mk8. Various rock classification methods, including seismic velocity-based, graphical, and grading approaches, have been developed to characterize lithologies based on rock mass strength (Hoek and Brown 1997; Barton 2006; Caterpillar Incorporation 2010; Akingboye 2023), as illustrated in Fig. 15. These methods integrate laboratory and field measurements, correlating seismic velocities with lithological properties such as rippability, rock mass strength, and ratings, enabling real-time data evaluation (Gallardo and Meju 2007; Zeng et al. 2018; Doyoro et al. 2025). Table 1 provides velocity and resistivity values for common crustal materials and minerals.

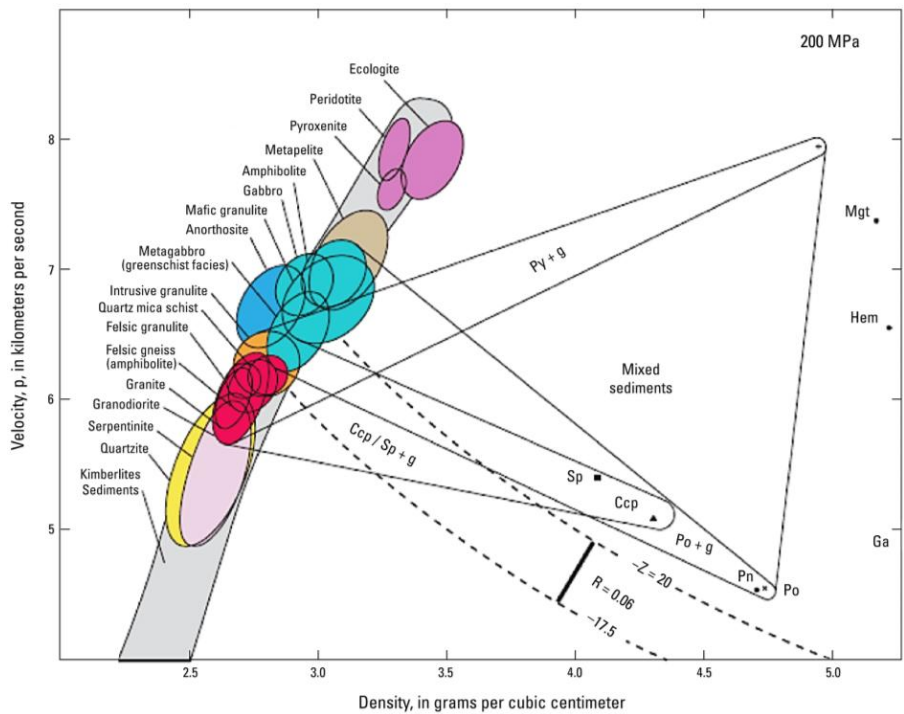

**Fig. 15.** Image of velocity (Vp) vs density values of various rock types and ore minerals. Lines of constant acoustic impedance (Z) are overlain within the field. *Pyrite (Py), pentlandite (Pn), pyrrhotite (Po), chalcopyrite (Ccp), sphalerite (Sp), hematite (Hem), and magnetite (Mgt), with other fields for host rock-ore mixtures. Galena (Ga), with a velocity of 3.7 km/s and a density of 7.5 g/cm³, is off the scale. A reflection coefficient (R) of 0.06 is indicated to produce a strong reflection in contrasting lithologies* (adapted from Morgan 2012).

Diverse analytical approaches, including qualitative, quantitative, statistical, and ML techniques, further refine classification methods, improving resolution in surface process studies and hazard assessments (Yang and Ma 2019; Li et al. 2020; Dick et al. 2025). SRT data effectively predict lithology, pore fluids, and rock structures, proving valuable for groundwater and mineral exploration, engineering site investigations, mining, and archaeology across various spatial scales (Zakaria et al. 2022; Sujitapan et al. 2024). It also resolves velocity gradients and lateral velocity variations within the subsurface. However, challenges persist in accurately delineating lithologic units due to masked layers, hidden layers, and water infills. Addressing these limitations requires incorporating additional rock properties and refining filtering techniques. Integrating SRT with complementary methods such as borehole logs, rock quality designation (RQD), and ERT enhances the reliability and completeness of subsurface geological models.

**Table 1.** Typical electrical P- and S-wave velocity and resistivity values for some selected rocks and other materials.

| Common Materials | P-wave velocity (m/s) | S-wave velocity (m/s) | Resistivity ($\Omega$m) |
|---|---|---|---|
| Air | 343 | N/A | Infinity ($\infty$) |
| Water | 1400 – 1600 (~1500) | N/A | 0.2 – 100 (up to 300 for exceptionally pure water) |
| Ice | 3400 – 3800 | 1700 – 1900 | 600 – $10^7$ |
| Oil | 1200 – 1400 | N/A | >200 |
| Clay | 900 – 2500 | 50 – 300 | 1 – 150 |
| Sand | 500 – 2200 | 100 – 800 | 10 – 1000 |
| Coal | 2200 – 2700 | 1000 – 1400 | 10 – 800 |
| Gravel | 500 – 2500 | 600 – 2250 | 100 – $10^4$ |
| Conglomerate | 1500 – 4500 | 1200 – 3000 | 2000 – $10^4$ |
| Limestone | 3500 – 6000 | 2000 – 3300 | 50 – 6000 |
| Shales | 1100 – 4500 | 200 – 800 | 20 – 2000 |
| Sandstone | 1400 – 4300 | 800 – 1800 | 10 – 5000 |
| Dolomite | 3500 – 6500 | 1900 – 3600 | 50 – 6000 |
| Granite | 4500 – 6000 | 2500 – 3300 | 300 (weathered) – $10^6$ |
| Basalt | 5000 – 6400 | 2400 – 2800 | 200 (weathered) – $10^6$ |
| Gneiss | 4400 – 5200 | 2700 – 3200 | 1000 – $3 \times 10^6$ |

*4.2 SRT field survey operations and data acquisition techniques*

SRT data acquisition has evolved with advancements in instrumentation, data processing algorithms, and survey design strategies. Modern SRT surveys utilize advanced seismic sources and receivers, including wireless systems, multichannel recorders, and three-component sensors, enhancing data quality and efficiency (Virieux and Operto 2009; Dahlin and Wisén 2018; Ronczka et al. 2018). The selection and deployment of seismic sources and receivers are critical to optimizing data quality. The ABEM Terraloc Mk8, for instance, is widely used (Fig. 16a). Common sources such as sledgehammers, explosive charges, and vibrators offer varying advantages in energy output, penetration depth, and safety considerations (Fang et al. 2020; Karslı et al. 2024). Receivers, typically geophones, are strategically positioned along survey lines to record seismic wave arrival times.

Survey design is essential for optimizing data acquisition, with geometries such as common offset and common midpoint used to ensure adequate coverage and resolution (Akingboye and Ogunyele 2019). During data acquisition, seismic waves generated by the source propagate through the subsurface and are recorded by receivers. Travel times are measured using a seismograph or data acquisition system (Fig. 16b). Key survey configuration settings include sampling interval, sampling number, and recording time. Proper verification of cables, geophones, and their sensitivities ensures reliable data acquisition (Akingboye 2023). SRT surveys are applicable in terrestrial (Sujitapan et al. 2024), crosshole (borehole) (Wong 1995), and marine environments (Schlindwein et al. 2003), with schematic layouts depicted in Fig. 16a–d. Recent advancements in instrumentation and computing have introduced innovative SRT data acquisition techniques. Distributed acoustic sensing (DAS) using fiber-optic cables enables high-resolution subsurface imaging with minimal surface disturbance over extensive survey areas (Virieux and Operto 2009; Fang et al. 2020). Furthermore, integrating SRT with complementary geophysical methods, including ERT, borehole logs, and GPR, enhances subsurface characterization and interpretation (Martinho and Dionísio 2014; Dick et al. 2024b).

**5. Electrical resistivity and seismic refraction tomographic data inversion and interpretation**

*5.1 Data processing, reduction, and inversion for ERT*

ERT data processing and inversion techniques transform raw field measurements into meaningful subsurface resistivity models. Preprocessing steps, including noise removal, quality control, and calibration, enhance data reliability (Loke 2002; Abdullah et al. 2019; Loke et al. 2022). Inversion algorithms then reconstruct resistivity distributions using software like RES2DINV, which supports least-squares inversion, Occam inversion, Tikhonov regularization, and Bayesian inversion (DeGroot-Hedlin and Constable 1990; Loke 2002; Bouchedda et al. 2017; Loke et al. 2022). These methods seek the most probable resistivity model while ensuring smoothness, geological consistency, and physical realism. The inversion program accommodates both conventional and non-conventional electrode arrays, including underwater and crosshole surveys, demonstrating adaptability across diverse conditions (Karaoulis et al. 2014; Abdullah et al. 2019).

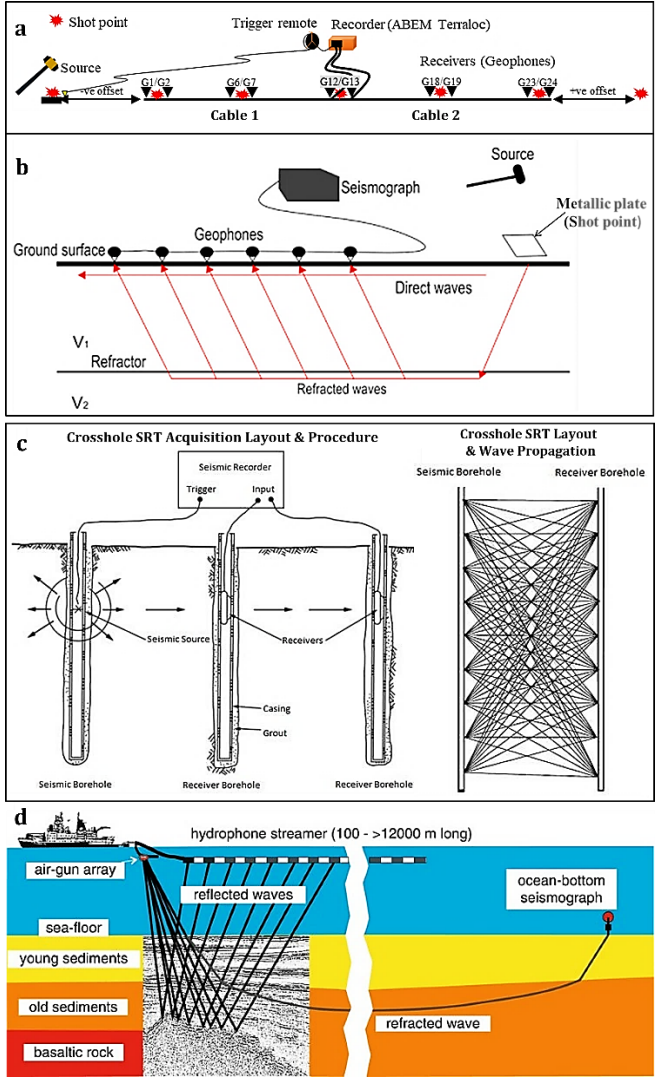

**Fig. 16.** (a) A typical layout for SRT data acquisition on land and (b) the progression of seismic waves for a two-layered subsurface model in an SRT ground survey. (c) Crosshole SRT survey type and (d) marine SRT survey procedures and acquisition layouts (adapted from Grootes 2016).

During inversion, the program utilizes either standard or robust least-squares approaches to minimize data misfits. Standard least-squares inversion iteratively adjusts model parameters to reduce discrepancies between observed and predicted data but is sensitive to outliers and noise, potentially leading to inaccuracies. Robust least-squares inversion, on the other hand, inversion mitigates this by assigning lower weights to data points with large residuals, ensuring greater stability and reliability, particularly in noisy datasets (Loke 2002). While standard least-squares inversion is computationally efficient, it can produce biased results in the presence of outliers. In contrast, robust least-squares, though more resilient, require greater computational resources and careful parameter tuning (Loke 2002; Akingboye et al. 2022). The choice between these methods depends on dataset characteristics, including noise levels and data quality. A key distinction lies in their misfit functions—standard inversion uses RMS error, while robust inversion employs absolute error for convergence evaluation.

ERT inversion solves the mathematical inverse problem by estimating subsurface resistivity from apparent resistivity measurements, iterating until the model best matches field data. The subsurface is discretized into rectangular cells, with their resistivities determined through inversion algorithms (Loke 2002). However, simplifications in the cell-based approach may not always yield true subsurface conditions. The linearized smoothness-constrained least-squares optimization method is described by DeGroot-Hedlin and Constable (1990) and Dahlin and Loke (2018), relating model parameters (r) to data misfit (g) as outlined in Eq. 23:

$$[J_i^T R_d J_i + \lambda_i W^T R_m W]\Delta r_i = J_i^T R_d g_i - \lambda_i W^T R_m W r_{i-1} \quad (23)$$

$J$ is the Jacobian matrix, representing apparent resistivity measurements relative to model resistivity values; $W$ is the roughness filter; and $\lambda$ signifies the damping factor. The data misfit vector, $g$, reflects the variance between measured and calculated resistivity values. $\Delta r_i$ denotes the necessary change in model parameters for iterations aimed at minimizing data misfits. $r_{i-1}$ is the resistivity model from the previous iteration, with $R_d$ and $R_m$ representing the weighting matrices used in the $L_2$-norm standard or robust inversion methods.

According to Cheng et al. (2019) and Akingboye and Bery (2021a), accurately assessing the root mean square error (RMSE) is crucial for quantitatively evaluating the soil-rock model. While synthetic models typically lack measurement errors, real field models are prone to such errors, which can significantly impact computed resistivity models. The RMSE, indicating the spatial variability of resistivity values, is computed as the disparity between measured and calculated resistivity values, following Eq. 24 (Loke 2002).

$$RMSE = \sqrt{\frac{1}{n^2}\sum_{i=1}^{n}\left(\frac{\rho_{mi} - \rho_{ci}}{\rho_{ci}}\right)} \quad (24)$$

Where $n$ is the number of measurement points, $\rho_{mi}$ and $\rho_{ci}$ are the measured and calculated resistivity and $i^{th}$ data point.

Recent advancements in ERT data processing include the integration of ML and AI, which offer the potential to improve inversion accuracy and efficiency. ML algorithms, such as ANNs, CNNs, Monte Carlo, GANs, ensemble learning, etc., have shown promising results in automating the inversion process and handling complex geological structures (Aleardi et al. 2021; Hiskiawan et al. 2023). Additionally, parallel and distributed computing architectures accelerate large-scale ERT dataset processing, while advanced regularization techniques and prior geological or geophysical constraints further enhance model robustness (Liu et al. 2020; Abdullah et al. 2022). Despite these advancements, challenges persist in accurately modeling complex geological structures, addressing non-uniqueness in inversion solutions, and effectively integrating multi-scale and multi-physics data. Overcoming these limitations requires interdisciplinary collaboration between geophysicists, mathematicians, computer scientists, and domain experts to develop innovative solutions that push the boundaries of ERT data processing and inversion techniques.

*5.2 Data processing, reduction, and inversion for SRT*

SRT data processing extracts meaningful subsurface insights through key steps, including preprocessing (filtering, noise removal, and time–distance corrections) to enhance signal quality (Dick et al. 2024b; Penta de Peppo et al. 2024). Arrival times are converted into velocity models using traveltime tomography algorithms, with inversion methods such as first-arrival traveltime tomography, waveform tomography, and full-waveform inversion balancing computational complexity and model resolution (Yang and Ma 2019; Fang et al. 2020). Filtering enhances signal-to-noise ratios, while arrival time picking identifies first-arrival waves crucial for velocity analysis.

Seismic inversion reconstructs subsurface velocity models by iteratively optimizing parameters through tomographic methods and iterative solvers (Côrte et al. 2020). Unlike conventional seismic refraction, which segments the subsurface into continuous velocity layers, SRT employs a grid-based approach for finer representation (Akingboye and Ogunyele 2019). Iterative ray tracing refines velocity models by comparing observed and calculated travel times until convergence (Salleh et al. 2021; Akingboye 2023). Regularization ensures solution stability, mitigating the ill-posed nature of the inverse problem. The final stage generates 2-D/3-D visualizations, aiding interpretation. Emerging 4-D SRT modeling is advancing subsurface monitoring, akin to TLERT (Côrte et al. 2020; Rosa et al. 2020).

Case studies across diverse geological settings demonstrate SRT's effectiveness in characterizing subsurface structures, detecting geological anomalies, and delineating geological features (Côrte et al. 2020; Adeola et al. 2022; Sujitapan et al. 2024; Doyoro et al. 2025). Advances in computational techniques, including parallel computing ML have improved the speed and accuracy of SRT inversions (Yang and Ma 2019; Liu et al. 2021). Despite these advancements, challenges remain, such as handling noisy data, resolving non-uniqueness, and integrating prior geological information. Future research should focus on developing advanced inversion algorithms for large-scale datasets, integrating multi-physics data, and improving uncertainty quantification methods.

*5.3 Interpretation: machine learning (ML) prospects in ERT and SRT*

As geophysical exploration grows in complexity, accurately estimating resistivity and velocity distributions becomes increasingly challenging. ML techniques have revolutionized ERT and SRT inversions, enhancing lithological interpretations (Yang and Ma 2019; Liu et al. 2020; Ibrahim et al. 2022; Alam et al. 2025). Algorithms such as ANNs, CNNs, RF, decision tree (DT), GBoost, CatBoost, etc., improve subsurface analysis, inferring resistivity and velocity structures, identifying seismic phases, and delineating geological interfaces (Li et al. 2020; Liu et al. 2020; Aleardi et al. 2021; Hiskiawan et al. 2023; Kundu et al. 2024; Doyoro et al. 2025). These techniques efficiently interpret large-scale ERT datasets and address nonlinearities in resistivity and seismic Vp data analysis, supporting studies on geological hazards, subsurface dissolution, groundwater migration, and lithological differentiation (Ibrahim et al. 2022; Dimech et al. 2022; Xixi et al. 2023; Sujitapan et al. 2024; Dick et al. 2025).

Recent research, like Liu et al. (2020) developed ERSInvNet, a CNN-based model using U-Net architecture, for real-time mapping of resistivity structures, incorporating depth weighting and smooth constraints to enhance inversion accuracy. Liu et al. (2021) proposed a deep-learning-based full-waveform inversion for Vp models, improving structural mapping of dense layers, faults, and salt bodies. Li et al. (2020) introduced SeisInvNets, a deep neural network architecture that reconstructs velocity models from seismic data, addressing nonlinear mapping and non-uniqueness issues. Although these methods improve inversion accuracy, generalizing deep-learning approaches to real data remains challenging. Incorporating physics-based constraints may enhance their robustness and applicability. Additionally, unsupervised ML algorithms, including k-means, hierarchical density-based, and fuzzy-based clustering, effectively identify structural patterns in geological formations, facilitating ERT and SRT data interpretation (Di Giuseppe et al. 2014; Bernardinetti and Bruno 2019; Delforge et al. 2021). Supervised learning techniques, such as statistical regressions, quantify relationships between resistivity, velocity, and geological parameters, enabling lithology, porosity, and fluid content estimation (Gallardo and Meju 2007; Zeng et al. 2018; Akingboye and Bery 2023a). Also, neural networks predict resistivity and velocity profiles, capturing complex variations to characterize geological units, fault zones, and stratigraphic boundaries (Latrach et al. 2024).

Efforts to establish velocity–resistivity correlations have advanced subsurface characterization. Early studies (Weatherby and Faust 1935; Faust 1951, 1953) demonstrated that seismic velocity increases with geological age and depth, leading to empirical relations correlating velocity with resistivity. Subsequent works improved these relationships using borehole data (Acheson 1963; Kim 1964). Rudman et al. (1975) demonstrated that conventional resistivity logs could generate pseudo-velocity logs, improving seismic velocity interpretation. More recent studies have enhanced these relationships across diverse geological settings, as presented in Table 2. Despite advancements in supervised and unsupervised ML, challenges persist in accurately predicting subsurface properties and integrating domain knowledge, particularly in physics and engineering (Liu et al. 2020, 2021; Akingboye and Bery 2023a). Lithological characterization remains complex due to subsurface heterogeneity (Gallardo and Meju 2011). However, ongoing research aims to overcome these challenges through hybrid ML models that integrate domain expertise with data-driven approaches.

Given the above findings, a systematic methodological framework is essential for efficiently modeling ERT and SRT data, from acquisition to establishing robust resistivity–velocity relationships. This framework enables rapid, detailed estimations of subsurface properties such as lithology, porosity, and fluid content based on statistical relationships. Hence, the proposed study framework, outlined in Fig. 17, builds upon the methodologies of Akingboye and Bery (2023a). Figure 17a illustrates the overarching methodology for establishing velocity–resistivity relationships, which is further detailed in Figs. 17b and 18. Specifically, Fig. 17b focuses on simultaneous data inversions, while Fig. 18 outlines interface modeling of resistivity and Vp models using ML techniques (e.g., SLR, SVM, GBoost, ANN), alongside k-means clustering and graphically derived regression modeling. Additionally, Fig. 18 includes accuracy assessments of the empirical relationship and predictive model, employing statistical validation parameters such as histograms, R² values, normal P-P plots, scatterplots, Durbin-Watson statistics, and p-values. Despite the site-specific nature of empirical relationships, this cost-effective and rapid approach bridges data acquisition gaps across diverse geological terrains, with practical applications demonstrated in case studies.

**Table 2.** Summary of selected recent analytical modeling and interpretation techniques (including ML) used for Vp and resistivity (DC, ERT, or magnetotelluric datasets) analysis.

| Location (or terrain) | Methodology | Proposed statistical/ empirical relationships | Accuracy assessment | References |
|---|---|---|---|---|
| Mountsorrel granodiorite terrain (characterized by a buried hillside of highly fractured granodiorite rock mass successively overlain by a heterogeneous mudstone and glacial drift) | Conjugate gradient inversion and statistics (regression) | $\text{Log}_{10}\,\rho = 3.88\,\text{Log}_{10}\,Vp - 11$ (for consolidated rocks at >3 m depth)<br>$\text{Log}_{10}\,\rho = -3.88\,\text{Log}_{10}\,Vp + 13$ (unconsolidated soil/drift deposits)<br>$\text{Log}_{10}\,\rho_a = 3.88\,\text{Log}_{10}\,Vp + 3.88\,\text{Log}_{10}\,11$ (solid grains with Vp less than pore fluid) | Porosity was considered a relevant factor based on the 3rd relation | (Meju et al. 2003) |
|  | Cross gradient function based on Lagrange multiplier | ----- | ---- | (Gallardo and Meju 2003) |
|  | Joint inversion formulation using forward and derivative approaches | ----- | RMSE (data misfit) with convergence <2% | (Gallardo 2004; Gallardo and Meju 2007) |
| Sub-basalt (Ghawar field, Saudi Arabia) | Simultaneous joint inversion and statistics | $Log\,\rho = 0.000441244(Vp) - 3.33559$ | 3-D generalized cross-gradient function | (Colombo et al. 2008) |
|  | Cross-plot regression function | $\rho = aV_p^{0.25}$ (the derived relation agrees with the provided Gardner's relation but with a slight coefficient change) | Confirmation of Gardner's relation | (Colombo and Keho 2010) |
|  | Joint wavefield inversion and Cross-plot regression function | $\rho = (3 \times 10^{-17})V_p^{5.0613}$ | $R^2 = 0.9127$ | (Colombo et al. 2020) |
| Groß Schönebeck Geothermal Field in Berlin, Germany | Gaussian probability density function | $f(x) = \sum_{j}^{n} \frac{a_j}{2\pi[\Sigma_j]^{1/2}} exp -\frac{1}{2}[(x - \varphi_j)^T \sum_{j}^{-1}(x - \varphi_j)]$ | Utilized the codes provided by Becken and Burkhardt (2004) | (Muñoz et al. 2010) |
| Swelling mudstone (Yanji Basin, NE China) | Regression statistics | $\rho = 133.07 - 0.074Vp$ (wetting–drying cycles)<br>$\rho = 13.525 + 0.023Vp$ (freezing–thawing cycles)<br>$\rho = 27.912 + 0.012Vp$ (wetting–drying–freezing–thawing cycles) | $R^2 = 0.816$<br>$R^2 = 0.752$<br>$R^2 = 0.641$ | (Zeng et al. 2018) |
| Granitic terrain (Penang Island, Malaysia) | Complex collocated geotomographic and supervised statistical modeling | $Vp = 0.4865(\rho) + 168.764$ (Generalized relation)<br>$Vp_{RS} = 0.1927(\rho_{RS}) + 363.32$ (residual soil unit)<br>$Vp_{W/WF} = 0.4713(\rho_{W/WF}) + 249.28$ (weathered/fractured unit)<br>$Vp_{I/FB} = 0.2506(\rho_{I/FB}) + 1193.9$ (integral/fresh granitic bedrock) | $R^2 = 0.8517$<br>$R^2 = 0.99$<br>$R^2 = 0.98$<br>$R^2 = 0.99$<br><br>Other statistical charts | (Akingboye and Bery 2023a) |
| Sedimentary terrain (Kabota-Tawau area of Sabah, Malaysia) | Interpolating P-wave velocity and resistivity models and machine learning (K-means clustering and supervised statistical modeling) | $\rho = 0.053(Vp) - 20.31$ (Generalized Vp and resistivity relation for surface–subsurface lithological prediction and modeling) | $R^2 = 0.86.8$<br><br>Different statistical and cluster density plots | (Dick et al. 2024b) |

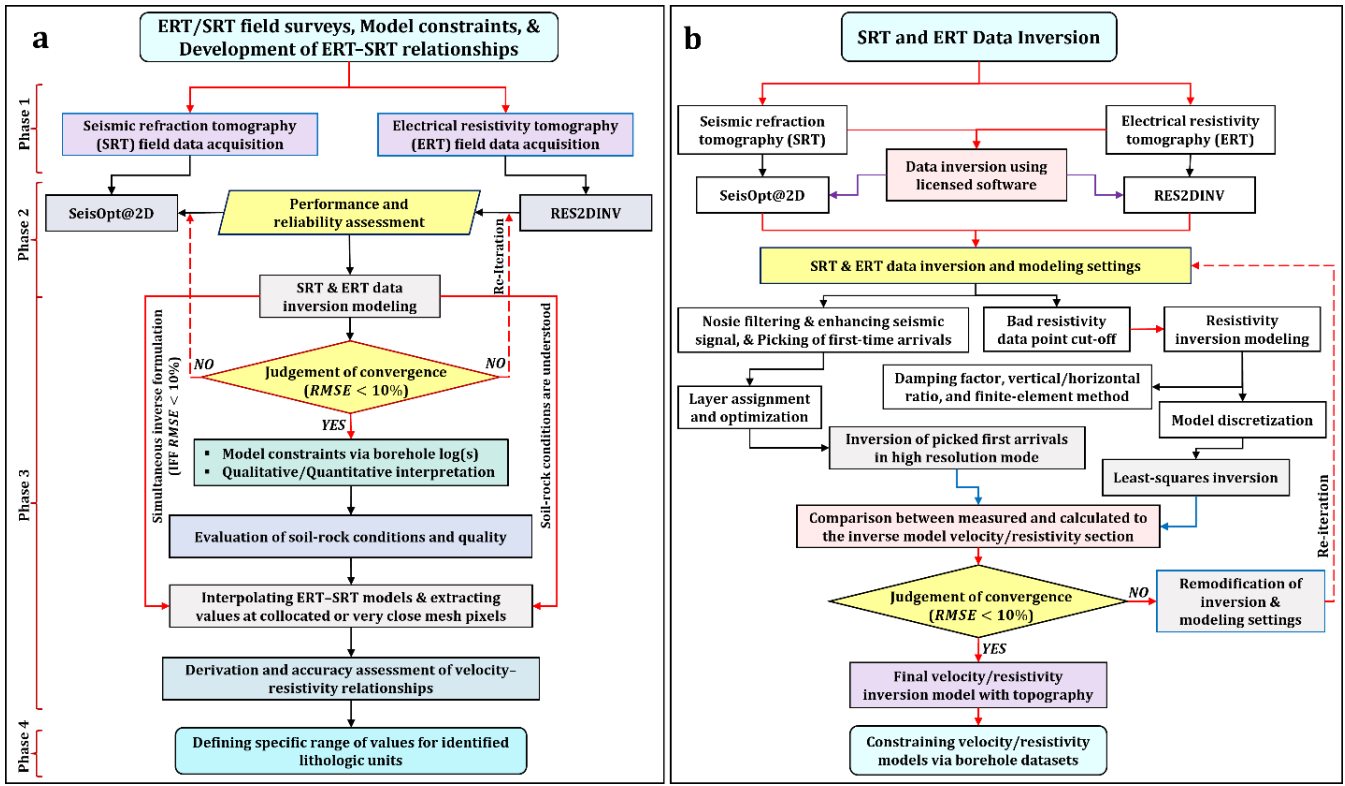

**Fig. 17.** (a) Proposed methodological framework for acquisition to supervised statistical modeling of ERT–SRT datasets. (b) Simultaneous processing and inversion of ERT and SRT datasets.

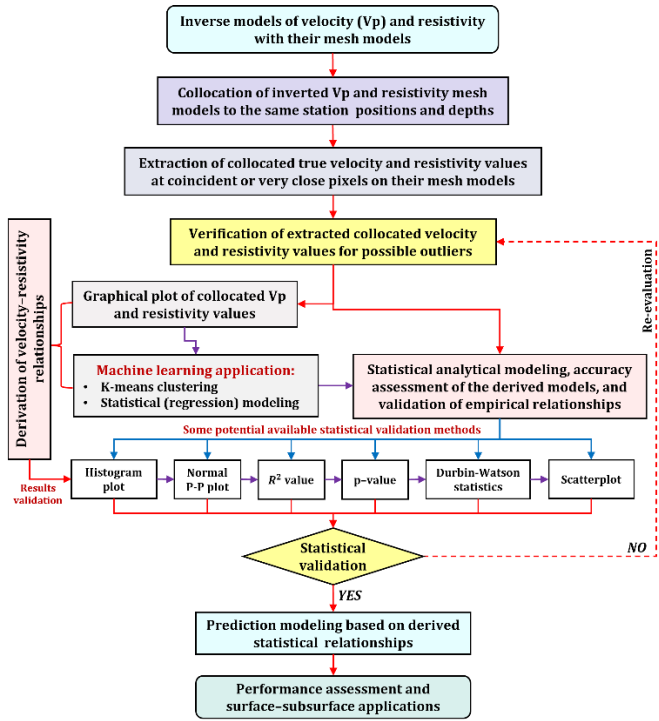

**Fig. 18.** Proposed framework architecture for supervised statistical modeling (including K-means clustering) of interfaced resistivity and velocity datasets.

### 5.4 Interpretation: cross-plot method for resistivity and velocity modeling

Cross-plot analysis is another well-established approach for interpreting velocity and resistivity datasets, widely applied in geosciences, including the oil and gas industry (Anyiam et al. 2018) and engineering studies such as landslide assessments (Zakaria et al. 2021; Sujitapan et al. 2024). This method involves plotting resistivity against seismic Vp or Vs (Fig. 19) to classify soil types and assess subsurface lithological conditions based on quadrant distributions (Hayashi and Konishi 2010). Intrusive borehole sampling is often integrated with cross-plotting to calibrate and validate geophysical data, improving accuracy in landslide assessments and mitigation planning. For instance, in Fig. 19a, the cross-plot of resistivity and Vp in the Q1 section identifies lithological materials susceptible to sliding, particularly in landslide-prone areas. Additionally, the other sections (Q2, Q3, and Q4) illustrate the variability in the degree of saturation, porosity, clay content, and compactness of crustal materials, which can influence various fields of geophysical interpretive modeling.

Hayashi and Konishi (2010) demonstrated the effectiveness of cross-plot analysis for levee assessments, specifically the relationship between resistivity and Vs, as shown in Fig. 19b. Vs indicates shear stiffness or the degree of compaction under constant effective stress, while resistivity reflects soil type, including grain size distribution and clay content. The diagonal vulnerability arrow in Fig. 19b illustrates the schematic relationship between Vs and resistivity and their influence on levee stability, indicating that loose, sandy levees present greater risks than compact, clayey levees. Building on the cross-plot model approach, Sujitapan et al. (2024) utilized cross-plot analysis to enhance subsurface resolution by integrating resistivity and Vp datasets for landslide detection in the Thungsong district, Nakhon Si Thammarat, Thailand. Their 2-D cross-plot model provided detailed subsurface images, identifying areas with low Vp and low resistivity as potential sliding zones. Similarly, Zakaria et al. (2021) applied cross-plot modeling of Vp and resistivity data in Ulu Yam, eastern Kuala Lumpur, Malaysia, to investigate landslide activity.

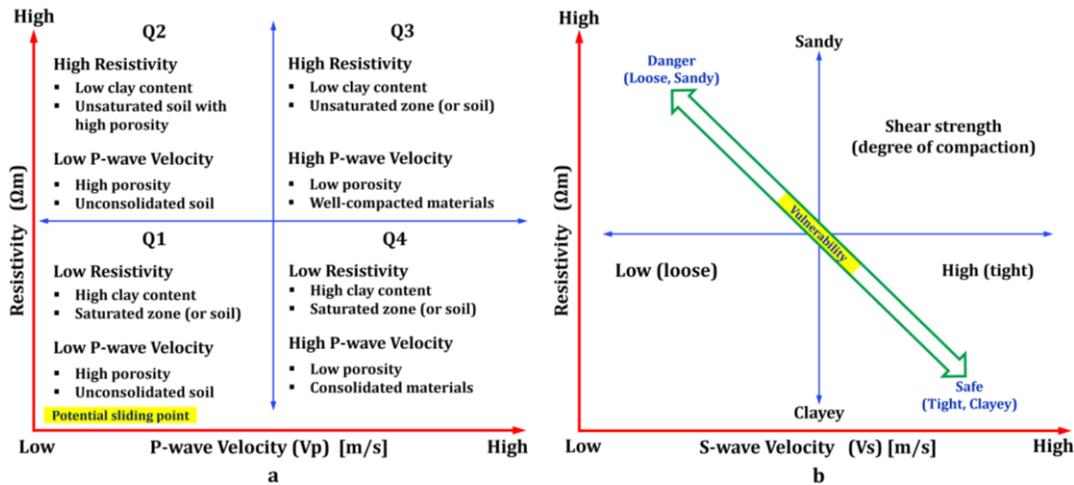

**Fig. 19.** Schematic cross-plots of resistivity against (a) P-wave velocity (Vp) and (b) S-wave velocity (Vs) for interpretation of soil-rock conditions (modified after Hayashi and Konishi 2010; Zakaria et al. 2021). The potential vulnerability of a levee channel is also shown by the Vs plot.

## 6. Case studies

The case studies further provide detailed insights into methodological approaches, 1-D/2-D qualitative and quantitative interpretations, and ML-derived analyses of resistivity and seismic Vp data/models across diverse geological terrains. The data acquisition and inversion methods for both ERT and SRT datasets in the case studies follow the methodologies previously outlined. For ERT, field data was collected using ABEM SAS Terrameter 4000 and ES 64–10C instruments. Resistivity datasets, along with topographical information, were processed and inverted using RES2DINV software. Depending on data noise levels, both standard and robust least-squares inversion techniques were applied. A finite-element method with constraint parameters, including a damping factor of 0.05 (minimum 0.01) was applied to enhance model resolution and to ensure that RMS or absolute errors of the inverse models converged below 10% within 5 to 7 iterations.

For SRT, data was collected using ABEM Terraloc Mk8, a 12-pound hammer, a striker metal plate, and a 24-channel geophone setup at 14 Hz. Instrument settings included a sampling interval of 100, 2048 samples, and a record time of 204.8 ms. Multiple shots (5 to 7 per shot point) ensured high-resolution models with increased depth penetration. Additional shots were deployed at offsets (10–12 m) to achieve deeper probing depths. Each SRT line recorded a minimum of 360 seismic traces from 7 to 15 shot points, depending on traverse length. Following field measurements, data processing involved managing observed, recorded, and source files, as well as shot point distances relative to station elevations. Weak signals were amplified using auto-gain control, and first arrivals were picked using FirstPix. SeisOpt@2D software (Optim Incorporation 2006), based on nonlinear optimization techniques, was then used for processing and inversion. The refraction inversion and optimization technique (RIOT) was applied, employing a fast finite-difference method to solve the eikonal equation for forward modeling and Vp dataset inversion. The resulting ERT and SRT models were constrained by lithologic units and depths obtained from borehole logs.

### 6.1 Case 1: electrodes' performance in surface–subsurface characterization

The study sites are located on Penang Island, Malaysia (Fig. 20), within the Malacca Strait, northwest of Peninsular Malaysia. The area is underlain by granitic rock, which is overlain by the Gula, Beruas, and Simpang Formations (Hassan 1990; Ong 1993) (Fig. 20a). The granite, similar to other Malaysian granitoids, is divided into two granitic plutons—the North Penang Pluton (NPP) and the South Penang Pluton (SPP), classified based on the alkali-to-total feldspar content (Ong 1993). The NPP includes the Tanjung Bungah, Paya Terubong, and Batu Ferringhi groups, along with the Muka Head microgranite, with ages spanning from the Early Permian to the Late Carboniferous period. In contrast, the SPP is distinguished by the Sungai Ara and Batu Maung granites (Ong 1993).

To evaluate the performance of copper and stainless-steel electrodes, a site within Universiti Sains Malaysia (USM), designated as Site 1, was investigated (Fig. 20b). Traverse S1 extended 200 m, with 5 m electrode spacing. A total of 41 copper and stainless-steel electrodes were deployed simultaneously. The Wenner-Schlumberger array was employed for ERT surveys at both Sites 1 and 2, and the resulting field data were processed based on the outlined standard least-squares inversion technique. After filtering errors related to cultural/self-potential noise, results indicated that the stainless-steel electrode type had higher RMSEs (7.2% and 3.5% for resistivity and chargeability models, respectively; Fig. 21a–b) compared to the copper electrode type (7.0% and 1.4%, respectively; Fig. 21c–d). This discrepancy aligns with the higher conductivity of copper compared to stainless-steel electrodes (LaBrecque and Daily 2008).

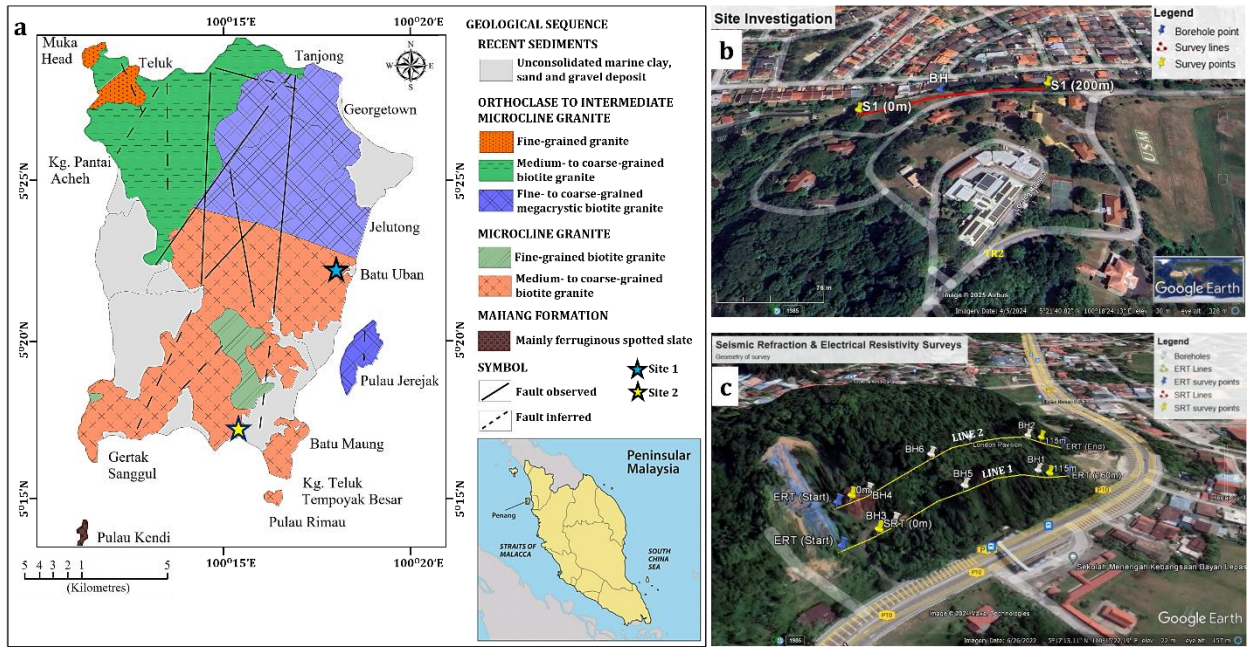

**Fig. 20.** (a) Penang Island's geological map (modified after Ong 1993), showing case study Sites 1 and 2. Aerial geophysical data acquisition maps of (b) Site 1 (at USM) and (c) Site 2 at Batu Maung area of Penang Island, Malaysia.

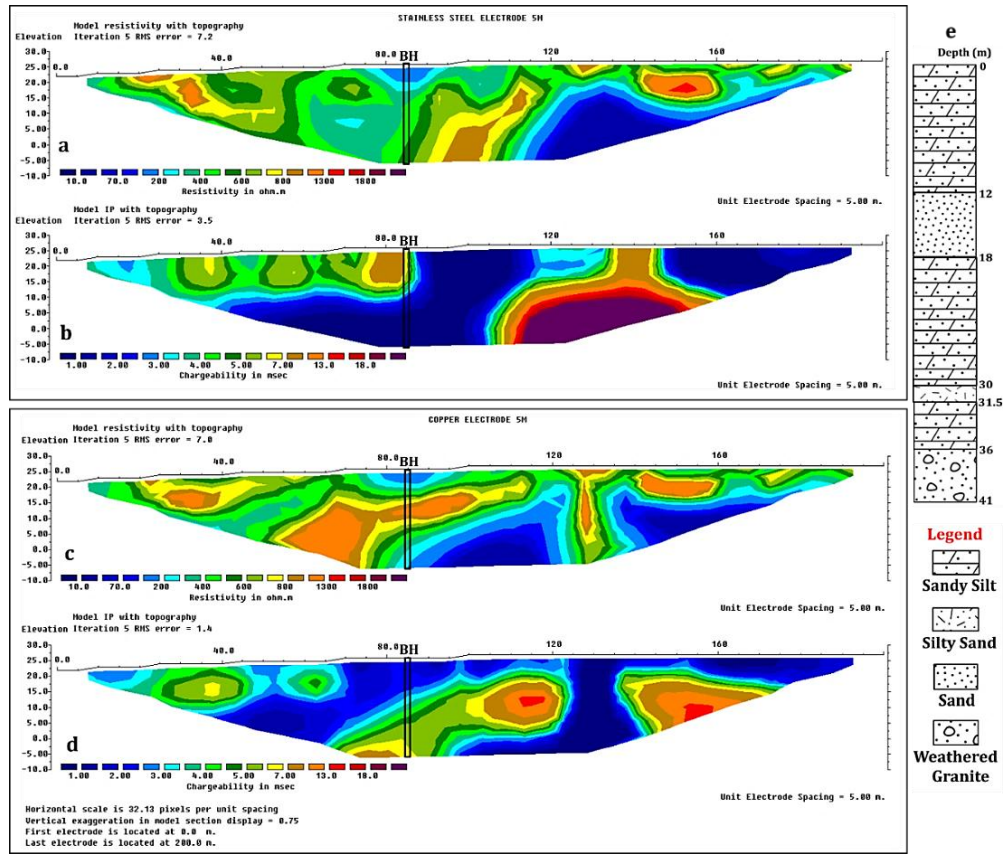

**Fig. 21.** ERT and IP inversion models of Site 1 (S1) generated using (a–b) stainless-steel electrodes and (c–d) copper electrodes, both with 5 m electrode spacing, along with (e) the borehole litholog within USM.

Both electrode types exhibited similar lithological features in the uppermost section, particularly at depths shallower than 6 m. However, discrepancies emerged at greater depths. Fig. 21c offers a more reliable geological representation than Fig. 21a, particularly in the central section (80–110 m station positions), where weathered granitic bedrock is observed (Fig. 21e). This feature was distinctly captured in the copper chargeability model (Fig. 21d), showing low to high chargeability, whereas Fig. 21b (stainless-steel model) exhibited uniformly low chargeability (<3 msec). Both resistivity and chargeability models effectively mapped the sandy silt layer between 110–160 m, characterized by very high chargeability (>10 msec). The obtained results corroborate the reviews on the sensitivity of both electrode types and their effectiveness in achieving high-resolution subsurface characterization.

## 6.2 Case 2: ML modeling of surface–subsurface conditions at Southern Penang Island, Malaysia

The study area in Batu Maung, southern Penang Island, Malaysia, designated as Site 2, serves as the focus for this investigation into subsurface lithological differentiation within a tropical granite terrain. To assess the distribution of seismic Vp and resistivity across surface–subsurface lithologic units, two geophysical survey lines were established in a SW–NE orientation (Fig. 20c). Each SRT line spans 115 m with 5 m geophone spacing, while ERT utilizes a 4.0 m electrode spacing over a traverse length of 160 m. Additionally, six boreholes (BH1–BH6) were drilled to constrain subsurface lithologic interpretations. The acquired datasets were processed using the aforementioned inversion procedures, producing subsurface Vp and resistivity models (Fig. 22a–e). From these models and borehole correlations, three distinct lithologic units were identified: clayey to sandy residual profile, weathered/fractured unit, and integral/fresh bedrock.

For ML modeling, resistivity and Vp inversion meshes were generated and interfaced, producing collocated models with perfectly aligned station distances and depths (Fig. 23a–b). A total of 110 datasets were extracted, ensuring near-perfect collocation, and analyzed using ML techniques, including SLR, SVM, GBoost, and ANN with a multilayer perceptron (MLP) architecture (as MLP-ANN), alongside k-means clustering and supervised regression statistical modeling. The ML models were employed to enhance the co-analysis of velocity–resistivity nonlinearity and improve predictive accuracy, with 70% (77) of the data used for training and 30% (33) for testing—an improvement that conventional inversion models struggle to achieve.

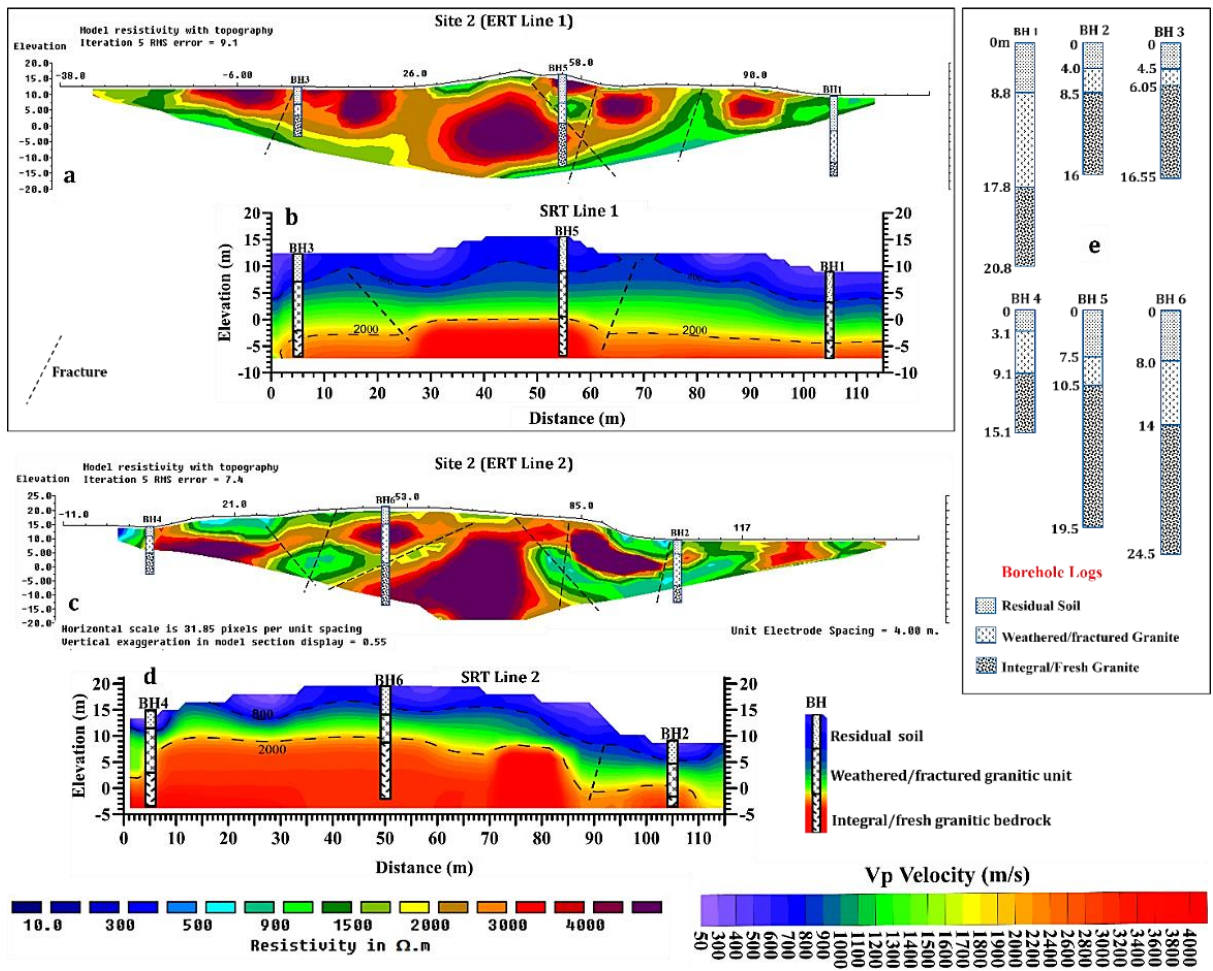

**Fig. 22.** 2-D ERT and SRT inversion models of (a–b) Line 1, (c–d) Line 2, and (e) the borehole lithological logs of BH1–BH6 at the Batu Maung area of Penang Island, Malaysia.

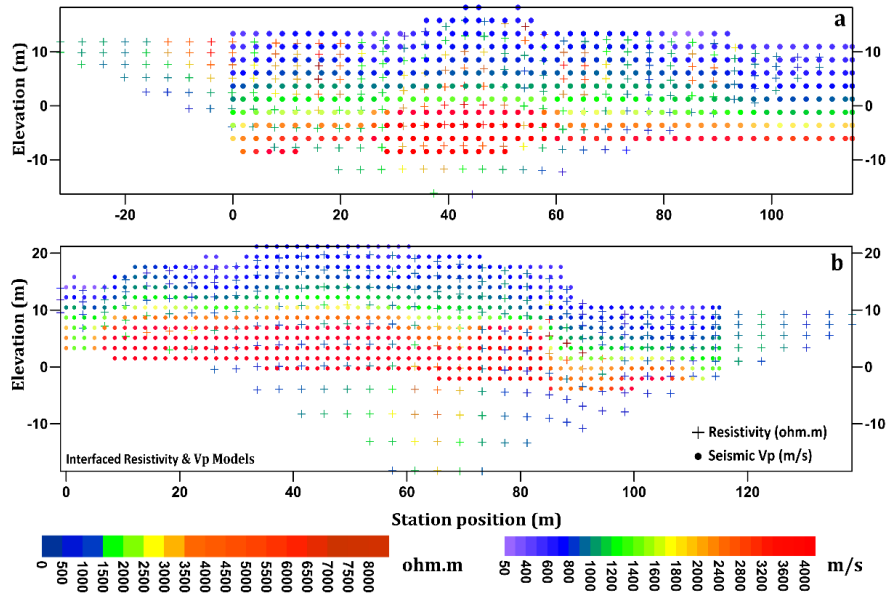

**Fig. 23.** Interfaced Vp and resistivity mesh models of section Lines 1 and 2 for Site 2 in the study area.

Furthermore, k-means clustering was applied for lithologic layer classification and to establish precise value ranges, leveraging the Elbow method—computed from the within-cluster sum of squares (WCSS)—and the average Silhouette scores to determine the optimal number of clusters (K) (Piegari et al. 2023; Doyoro et al. 2025). Originating from signal processing, k-means clustering is a vector quantization technique that iteratively assigns data points to the nearest cluster centroid and updates centroids based on the average of the assigned data points. This process continues until centroids stabilize or a predefined number of iterations is reached (Delforge et al. 2021). While k-means clustering effectively uncovers patterns in data, its performance depends on the optimal selection of cluster numbers and initialization methods (Di Giuseppe et al. 2014; Sinaga and Yang 2020).

Table 3 presents the performance metrics of ML models applied to resistivity and Vp datasets, evaluating both training and test sets using MAE, MSE, RMSE, $R^2$, and adjusted $R^2$. Among the models, GBoost demonstrated the highest predictive accuracy in the training phase, achieving the lowest MAE (0.048), MSE (0.005), and RMSE (0.072), alongside the highest $R^2$ (0.885). However, its test performance declined ($RMSE = 0.095, R^2 = 0.801$), indicating potential overfitting. In contrast, SLR exhibited a more consistent performance, with a lower training $R^2$ (0.843) but improved generalization, as evidenced by the lowest test MAE (0.060) and RMSE (0.078), along with a strong test $R^2$ (0.866). SVM displayed moderate predictive capability, with RMSE values of 0.080 (train) and 0.089 (test), while its test $R^2$ (0.824) remained lower than that of SLR and MLP-ANN. The MLP-ANN model showed the best generalization ability, achieving the lowest test RMSE (0.076) and highest $R^2$ (0.870), suggesting robustness in capturing complex nonlinear relationships. The adjusted $R^2$ values closely aligned with their $R^2$ values confirming minimal model complexity bias (underfitting/overfitting) (Akakuru et al. 2025). Overall, while GBoost excelled in training performance, MLP-ANN demonstrated the most stable generalization, making it the most reliable model for predicting velocity–resistivity relationships for the study area.

**Table 3.** Summarized metric performance for ML training and test sets for resistivity and Vp datasets ($N = 110$).

| ML models | Model sets | MAE | MSE | RMSE | $R^2$ | Adjusted $R^2$ |
|---|---|---|---|---|---|---|
| SLR | Train ($N = 77$) | 0.064 | 0.007 | 0.085 | 0.843 | 0.841 |
|  | Test ($N = 33$) | 0.060 | 0.006 | 0.078 | 0.866 |  |
| SVM | Train | 0.063 | 0.006 | 0.080 | 0.858 | 0.856 |
|  | Test | 0.068 | 0.008 | 0.089 | 0.824 |  |
| GBoost | Train | 0.048 | 0.005 | 0.072 | 0.885 | 0.883 |
|  | Test | 0.067 | 0.009 | 0.095 | 0.801 |  |
| MLP-ANN | Train | 0.062 | 0.007 | 0.082 | 0.852 | 0.850 |
|  | Test | 0.058 | 0.006 | 0.076 | 0.870 |  |

As shown in Fig. 24a–b, detailed analysis reveals that the WCSS and average Silhouette values for the optimum $K = 4$ are 1.123 and 0.547, respectively, indicating the most preferred elbow point to accurately capture the study area's lithologic units. The derived k-means clustering model for the analyzed collocated resistivity and Vp datasets is shown in Fig. 24c, depicting the classified layer units represented by Clusters 1–4 with their centroids for $K = 4$. This model aligns well with the regression-based graphical (statistical) model in Fig. 24d, effectively capturing the four clusters with their specific value ranges: clayey to sandy residual soil (<2400 Ωm, <800 m/s; Cluster 1), weathered/fractured unit (800–3350 Ωm, 800–1900 m/s; Cluster 2), integral bedrock (>3350 Ωm, 1900–2400 m/s; Cluster 3), and fresh bedrock (>3350 Ωm, >2400 m/s; Cluster 4). Though resistivity values overlap, this is an inherent characteristic of lithological resistivities, particularly in weathered and fractured layers (Akingboye and Bery 2023a).

The clusters accurately represent the surface-to-subsurface lithologies and their associated Vp and resistivity values, offering insights into the lithological conditions of a typical granitic terrain in the tropics. The empirical relation generated exhibits a very strong positive correlation of 0.923, with a p-value (<0.0001), significantly below the 0.05 (5%) threshold for statistically valid modeling. The regression-based graphical statistical modeling $R^2$ value (0.852) (Fig. 24d) aligns with the MLP-ANN training result (Table 3), indicating that the derived velocity–resistivity relationship is effective for predictions. Additionally, the high MLP-ANN test $R^2$ value of 0.87 (87% prediction accuracy; Table 3) confirms its reliability in predicting Vp values over a large area (Dick et al. 2024b). This suggests that ML approaches are effective, scalable tools for lithological modeling, helping to reduce interpretation ambiguities and minimize data acquisition costs across extensive areas.

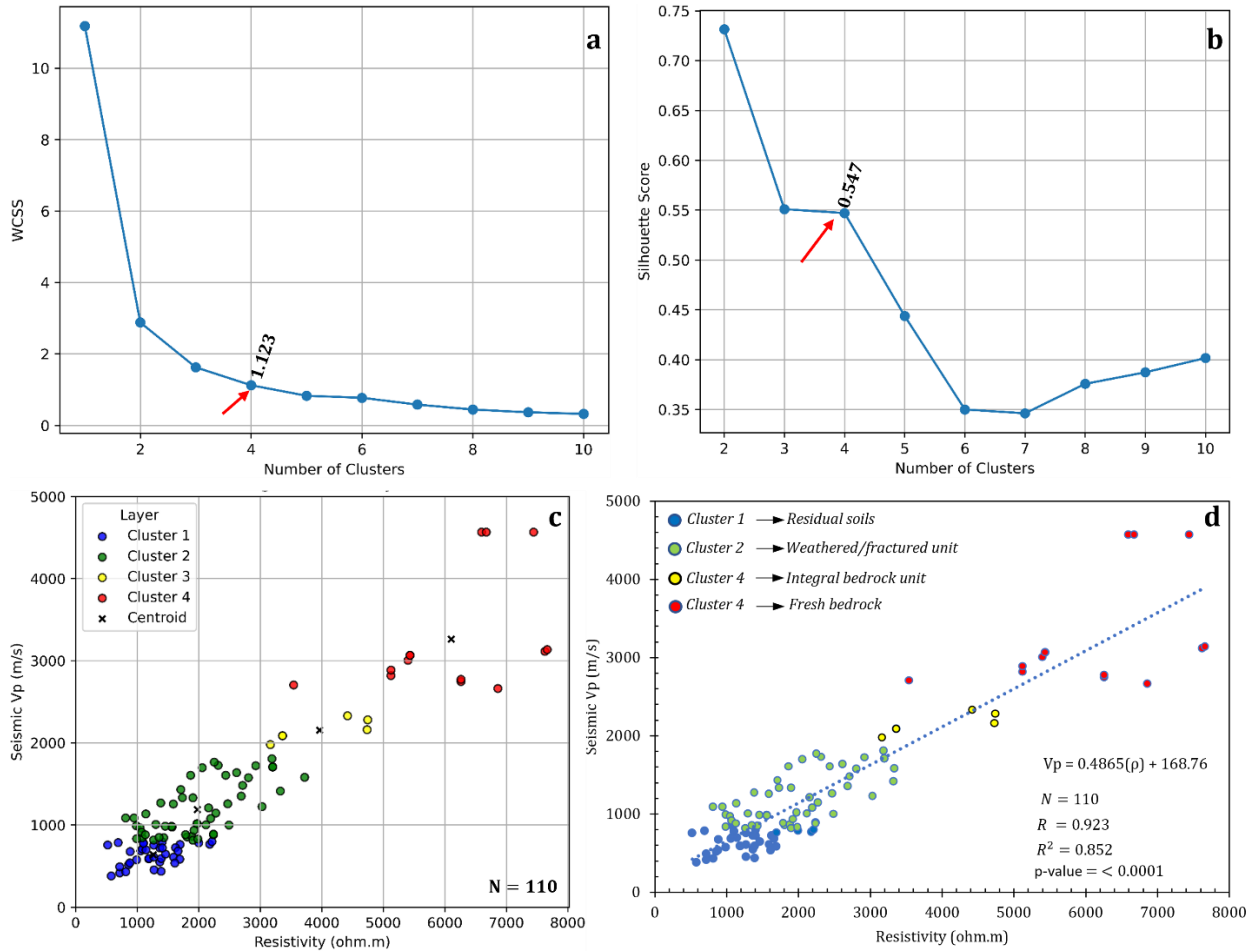

**Fig. 24.** ML-derived models: (a) Elbow method plot for optimal cluster determination, (b) Silhouette scores for various clusters, (c) k-means clustering results, and (d) regression-based graphical statistical modeling of velocity–resistivity relationships for the study area.

*6.3 Case 3: 1-D/2-D resistivity modeling of a metamorphic terrain in SW Nigeria—hydrogeological insights*

The case study is located in the Araromi area of Akungba-Akoko, Ondo State, southwestern Nigeria (Fig. 25a–d). This region falls within the Nigerian rainforest belt, characterized by distinct wet and dry seasons. Geologically, it lies within the Southwestern Nigerian Precambrian Basement Complex, a segment of the Pan-African mobile belt situated east of the West African Craton and northwest of the Congo-Gabon Craton (Woakes et al. 1987; Kröner et al. 2001; Obaje 2009; Oyeshomo et al. 2025) (Fig. 25a). The study area's underlying geology consists of the Migmatite-Gneiss Complex (MGC), intruded by Pan-African Granitoids (Ogunyele et al. 2020; Akingboye and Osazuwa 2021) (Fig. 25b). The MGC comprises migmatite, granite gneiss, and biotite gneiss, while the granitoids are predominantly charnockite and granite. Granite gneiss is the dominant lithology, with common intrusions such as quartz veins, pegmatites, aplite, basic dykes, and sills.

To assess the subsurface geological framework, six 2-D resistivity traverses (TRs) were conducted (Fig. 25c–d) using both dipole-dipole resistivity profiling and Schlumberger vertical electrical sounding (VES; 1-D resistivity imaging). TRs 1–3 and 4–6 were strategically positioned within a groundwater-deficient area due to complex subsurface geology. TRs 1, 2, and 6 extended 160 m, TR3 was 100 m, TR4 was 145 m, and TR5 covered 110 m. Traverse lengths varied due to infrastructural obstructions. VES surveys were conducted at select conductive stations, aiding in constraining ERT models in the absence of borehole data and delineating deep-weathered zones and fractures. ERT datasets were inverted using the outlined robust least-squares inversion procedures, while VES data were iteratively processed with IPI2win software to generate resistivity curve types with their layers' resistivities, thicknesses, and depths.

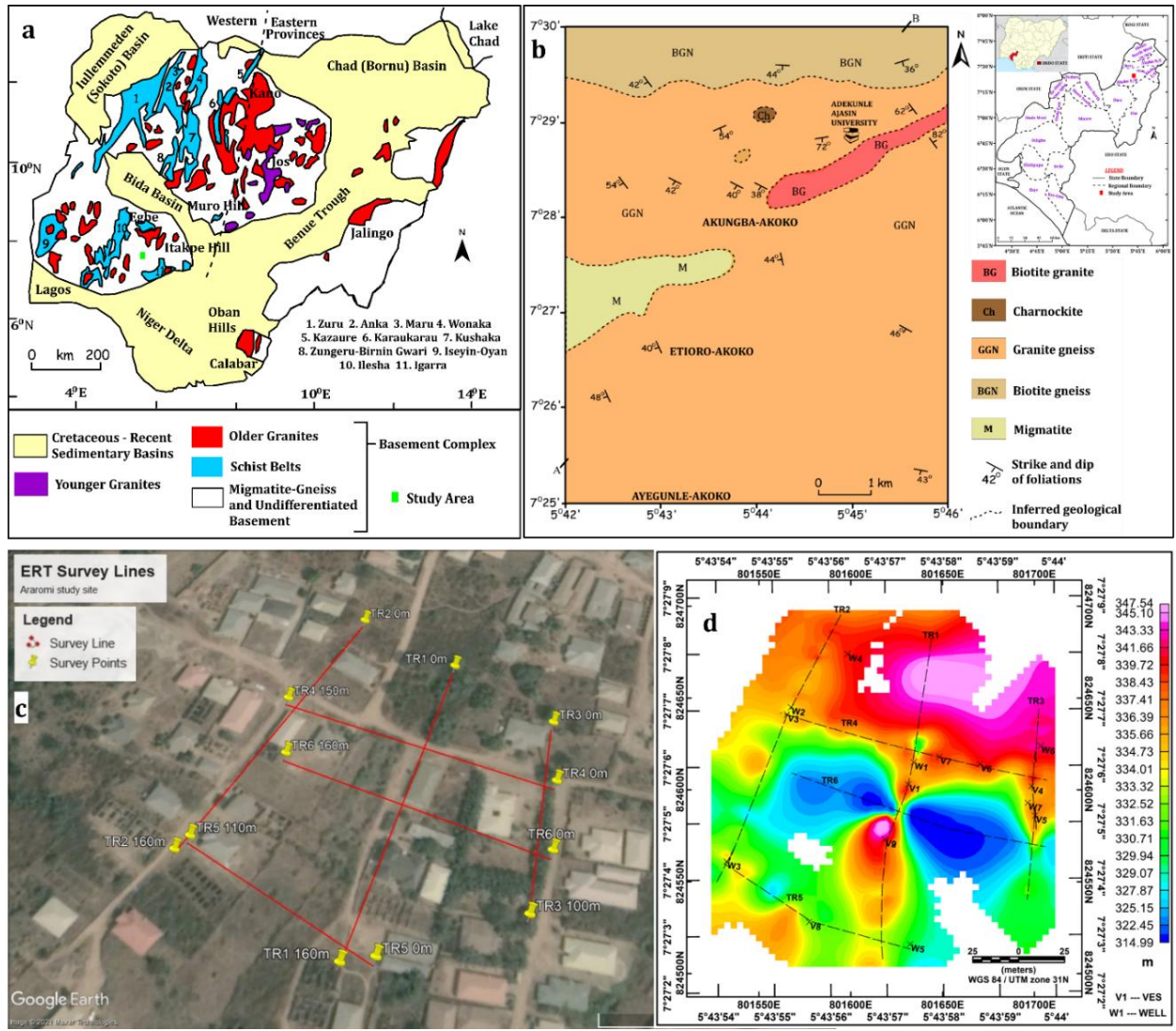

**Fig. 25.** (a) Nigerian regional geological map showing the study area within the Southwestern Basement Complex (modified after Woakes et al. 1987). (c) Geological map of Akungba-Akoko and its environs in Ondo State, SW Nigeria (after Ogunyele et al. 2020; Akingboye and Osazuwa 2021). (c) Aerial data acquisition and (d) elevation maps of the study showing all six ERT lines, eight VES survey points, and existing hand-dug well locations.

The ERT inversion models for the six traverses reveal four distinct subsurface layers: topsoil (<10–600 Ωm), weathered layer (600 to <1000 Ωm), weathered/fractured bedrock (>1200 Ωm), and fresh gneissic bedrock (>1200 Ωm) (Fig. 26a–f). Conductivity variations highlight the distribution of weathered materials, water-saturated zones, and sand-filled sections. VES results (Table 4) indicate topsoil thicknesses ranging from 1.16 to 3.8 m, while weathering extends to depths of ~10.3 m, influenced by fractures with variable-sized apertures. The weathered/fractured bedrock extends beyond 39 m at VES 7 (Fig. 26d; Table 4) and exceeds 52 m in certain sections (Table 4), suggesting potential groundwater zones. Fig. 26a–f further reveals intense deformation in the bedrock, likely due to high-grade metamorphism, resulting in undulating subsurface and localized open-to-surface fractures (Akingboye 2022). These deep-weathered sections, particularly along VES 7 and 8, present viable targets for sustainable groundwater abstraction in the region. These approaches have demonstrated effectiveness in groundwater exploration, particularly in complex geological terrains.

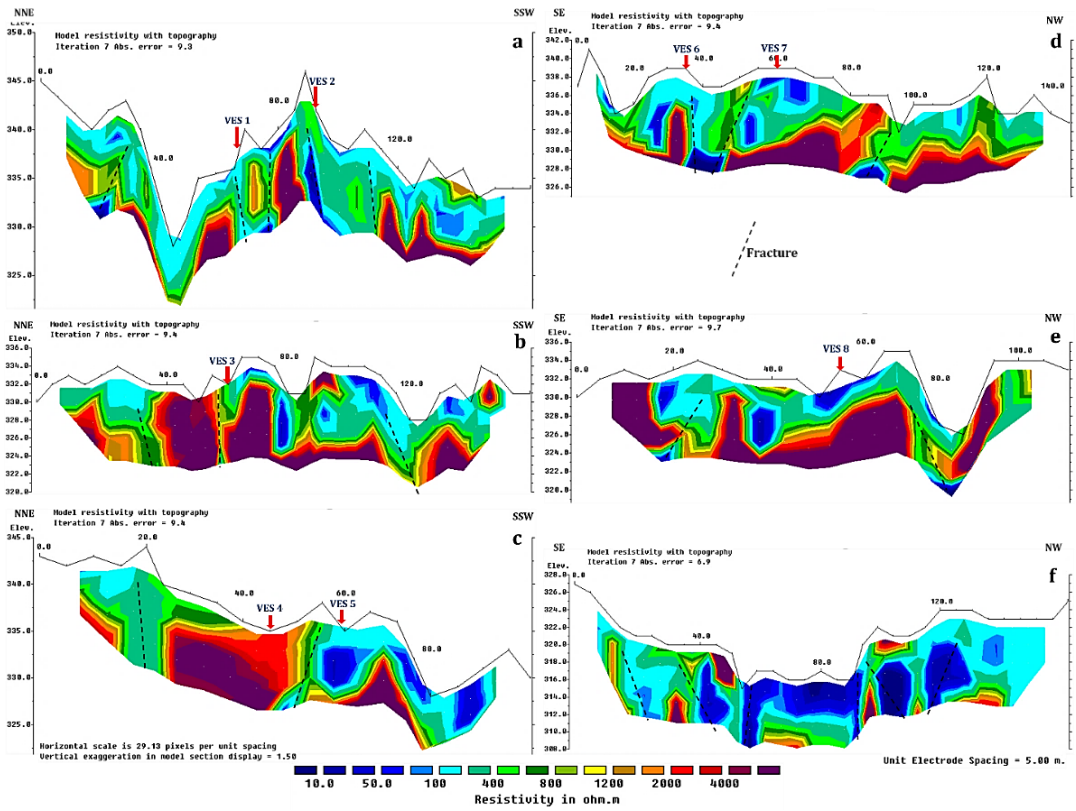

**Fig. 26.** ERT inversion models of (a) Traverse 1, (b) Traverse 2, (c) Traverse 3, (d) Traverse 4, (e) Traverse 5, and (f) Traverse 6. Red arrows show the VES point locations along the traverses.

**Table 4.** Iteratively generated VES curve type results with their inferred lithological interpretation.

| ERT Traverse | VES point | Station (m) | Curve type | Resistivity (Ωm) | Thickness, h (m) | Depth, H (m) | Lithological inferred interpretation |
|---|---|---|---|---|---|---|---|
| 1 | VES1 | 67.5 | KQ | 19.4 | 1.16 | 1.16 | Topsoil (clay-rich) |
| | | | | 7847 | 7.45 | 8.61 | Fresh gneissic bedrock slab |
| | | | | 415 | 43.60 | 52.20 | Deep weathered trough/fractured bedrock (water-saturated column) |
| | | | | 143 | ---- | | |
| | VES2 | 95 | HA | 110 | 1.51 | 1.51 | Topsoil |
| | | | | 29.2 | 3.13 | 4.64 | Water-saturated weathered trough |
| | | | | 699 | 50.10 | 54.80 | Fractured bedrock slab |
| | | | | 912 | ---- | | Partially weathered trough. |
| 2 | VES3 | 60 | A | 106 | 1.27 | 1.27 | Topsoil |
| | | | | 415 | 4.40 | 5.67 | Sandy weathered trough |
| | | | | 2736 | ---- | | Fresh gneissic bedrock |
| 3 | VES4 | 45 | A | 158 | 3.50 | 3.50 | Topsoil |
| | | | | 1218 | 12.7 | 16.20 | Gradually increasing resistive fresh bedrock slab |
| | | | | 5207 | ---- | | |
| | VES5 | 58 | A | 137 | 2.59 | 2.59 | Topsoil |
| | | | | 521 | 7.71 | 10.30 | Sandy weathered trough |
| | | | | 1288 | ---- | | Fresh bedrock slab |
| 4 | VES6 | 35 | A | 57.5 | 1.68 | 1.68 | Topsoil |
| | | | | 546 | 4.92 | 6.60 | Sandy weathered trough |
| | | | | 7451 | ---- | | Fresh bedrock slab |
| | VES7 | 60 | AK | 53.42 | 1.30 | 1.30 | Topsoil |
| | | | | 267.6 | 4.50 | 5.80 | Saturated sandy weathered trough |
| | | | | 2476 | 33.15 | 38.95 | Fresh bedrock slab |
| | | | | 838.1 | ---- | | Fractured bedrock column |
| 5 | VES8 | 55 | H | 567 | 1.35 | 1.35 | Topsoil |
| | | | | 138 | 5.17 | 6.52 | Sandy clay weathered trough |
| | | | | 3872 | ---- | | Fresh bedrock slab |

00

# 7. Conclusions

This comprehensive review delves into the intricacies of electrical and seismic refraction methods, including their emergent techniques—ERT and SRT—highlighting their critical roles in unraveling surface–subsurface crustal dynamics. Additionally, it explores the IP method and briefly touches on the SP method. From operational principles to data acquisition, processing, interpretation, and emerging prospects, this review provides in-depth insights into geophysical exploration. Case studies incorporating real-time resistivity and seismic Vp datasets further illustrate these concepts. ERT effectively images resistivity distributions, while SRT complements it by delineating subsurface velocity structures. Together, these methods are indispensable for hazard analysis, engineering applications, hydrological studies, environmental assessments, and resource evaluation. Their combined application enables comprehensive subsurface characterization, strengthening geological and geotechnical investigations. This review discusses their principles, acquisition techniques, processing methodologies, and interpretation strategies, underscoring both their strengths and limitations.

Beyond geophysical tools, this study highlights the transformative role of ML in refining resistivity and seismic Vp methods for subsurface characterization. The proposed methodological frameworks, supported by detailed case studies, offer a systematic approach to modeling ERT and SRT data for precise geological interpretation. The integration of ML techniques—SLR, SVM, GBoost, and ANN—alongside k-means clustering and statistical regression modeling has significantly improved the interpretation of collocated resistivity and Vp datasets. These advancements enhance velocity–resistivity modeling, improve inversion accuracy, automate lithological differentiation, and enable Vp prediction from resistivity data. The k-means clustering approach, optimized using the Elbow method and Silhouette analysis, effectively classifies lithological units, while regression-based ML models provide high predictive accuracy. Additionally, cross-plot analysis has proven valuable for interpreting velocity and resistivity datasets, aiding geological characterization and hazard assessments.

Despite these advancements, challenges remain. Lithological complexities, the nonlinear nature of geophysical properties, and uncertainties in subsurface data necessitate an integrated, multidisciplinary approach. Leveraging advanced ML techniques alongside domain-specific knowledge can reduce interpretation ambiguities and enhance subsurface modeling reliability. As demonstrated in case studies, ML-driven approaches streamline data analysis and minimize acquisition costs, making them scalable solutions for large-scale subsurface investigations. Their effectiveness in studied sites validates their applicability across diverse geological terrains, offering crucial insights for geotechnical assessments, geohazard mitigation, and sustainable groundwater development. Continued research in electrical and seismic velocity methods, coupled with advances in ML, will further refine geophysical modeling. Future developments in hybrid ML algorithms, improved inversion techniques, and physics-informed neural networks will enhance subsurface imaging and predictive modeling.

## Acknowledgments

The author expresses gratitude to Associate Prof. Dr. Andy A. Bery for his assistance throughout the collection of field data, as well as financial support contributed towards the use of data from Penang Island, Malaysia. Funds and state-of-the-art equipment provided by the Tertiary Education Trust Fund of Nigeria, Adekunle Ajasin University, Universiti Sains Malaysia (School of Physics), and Helmholtz Centre Potsdam – GFZ German Research Centre for Geosciences through HiDA Fellowship are greatly appreciated.

## Funding

This research is supported by funding from the Tertiary Education Trust Fund of Nigeria through Adekunle Ajasin University, Universiti Sains Malaysia through Dr. Andy A. Bery, and the PhD short-stay research grant from the Helmholtz Information and Data Science Academy (HiDA).

## Data Availability

All data generated or analyzed during this study are included in this published article. The author can make other supporting analyzed data available upon reasonable request.

## Declarations

**Ethical Approval:** All ethical standards have been duly followed during the research.

**Consent to Participate:** Not Applicable

**Consent to Publish:** Not Applicable

**Financial interests:** The authors declare no known competing financial interests or personal relationships that could have appeared to influence the work reported in this paper.